\documentclass[letterpaper]{JHEP3}
\usepackage{epsfig,verbatim}
\usepackage{subfigure}
\usepackage{amsmath, amssymb, graphics}

\newcommand{\mathsym}[1]{{}}

\newcommand{\eref}[1]{(\ref{#1})}

\renewcommand\({\left(}
\renewcommand\){\right)}
\renewcommand\[{\left[}
\renewcommand\]{\right]}

\newcommand{\dd}{{\rm d}}
\newcommand{\e}{{\rm e}}

\newcommand\eps{\epsilon}
\newcommand\mpl{m_{\rm p}}

\def\ba{\begin{eqnarray}}
\def\ea{\end{eqnarray}}
\def\be{\begin{equation}}
\def\ee{\end{equation}}

\def\L{\mathcal{L}}
\def\O{\mathcal{O}}
\def\P{\mathcal{P}}

\def\N{\mathcal{N}}

\def\nn{\nonumber}
\def\({\left(}
\def\){\right)}
\def\k{\kappa}
\def\Tr{{\rm Tr}}
\def\ini{{\rm in}}

\def\eref#1{(\ref{#1})}

\newcommand{\roughly}[1]{\mathrel{\raise.3ex\hbox{$#1$\kern-0.85em
\lower1ex\hbox{$\sim$}}}}

\title{M-flation and its spectators}

\author{M. Postma \\ Nikhef, Science Park 105 1098 XG Amsterdam, The
Netherlands. }

\date{}

\abstract {M-flation is an implementation of assisted inflation, in
which the inflaton fields are three $N_c \times N_c$ non-abelian
hermitean matrices.  The model can be consistently truncated to an
effectively single field inflation model, with all ``spectator''
fields fixed at the origin. We show that starting with random initial
conditions for all fields the truncated sector is {\it not} a
late-time attractor, but instead the system evolves towards quadratic
assisted inflation with all fields mass degenerate. Demanding the
energy density during inflation to be below the effective quantum
gravity scale, we find that the number of fields, and thus the
assisted effect, is bounded $N_c \lesssim 10^2$.  }

\preprint{NIKHEF 2010-026}

\begin{document}


\section{Introduction}

In chaotic models of inflation \cite{chaotic} the accelerated
expansion of the universe can be implemented using the simplest
potential one can think of, namely a single scalar field with a
polynomial potential.  The model predicts a large gravitational wave
signal, falsifiable by the upcoming Planck data. There is a price to
pay though. Inflation only works for very small couplings, for
example, the quartic self-coupling should be smaller than $\lambda
\lesssim 10^{-14}$.  Another, even larger, drawback of this type of
models is that the field value during inflation exceeds the Planck
scale.\footnote{It follows from the Lyth bound \cite{lythbound} that
in any single-field inflation model that predicts a measurable
gravitational wave spectrum, the effective inflaton field changes by
more than the Planck scale during inflation $\Delta \phi > \mpl$.} The
model is thus extremely UV sensitive: all non-perturbative operators
in the potential should be suppressed below their ``natural'' value to
retain perturbative control.
 
In assisted models \cite{assisted} inflation is driven by many scalar
fields simultaneously. This framework has the potential to address
both of the drawbacks of single field chaotic inflation: tiny
couplings and super-Planckian field values.  Assisted inflation may
proceed even if each of the individual fields has a potential too
steep for that field to sustain inflation on its own. This is because
each field experiences, via the altered expansion of the universe, the
frictional effect of all scalars. Since steeper potentials can lead to
inflation, larger couplings are allowed. Instead of having one field
with a large amplitude drive inflation, there are now many fields with
a small amplitude. This suggests that the UV behavior of the model is
under control \cite{assUV}.  However, this statement should be taken
with care \cite{Nflation}.  Perturbative and non-perturbative
arguments suggest that in theories with many species the effective
scale where quantum gravity becomes strong is below the Planck
scale~\cite{veneziano,dvali1,dvali2}.  If the field values are above
the effective quantum gravity cutoff during inflation, the problem of
UV sensitivity persists.

In most models of assisted inflation all scalars couple to each other
only gravitationally, as cross-couplings between the fields tend to
kill the assisted behavior \cite{Nflation,kanti}. The system can be
straightforwardly truncated to an effectively single field inflation
sector~\cite{assisted,malik}. Typically, this sector is a late time
attractor of the full system, and serves to describe the last 60
observable e-folds of inflation \cite{assisted,kanti}.
Although the couplings appearing in the effective single-field
potential are tiny, as needed for chaotic inflation, the couplings in
the original Lagrangian are not.  Likewise, although the effective
inflaton field has super-Planckian field values during inflation, the
individual fields appearing in the original Lagrangian do not.

Recently a matrix model of inflation was proposed, motivated from
string theory, in which inflation is driven by matrix-valued scalar
fields \cite{Mflation1,Mflation2}. M-flation, as this model was
dubbed, can be seen as a specific implementation of assisted
inflation. Truncating the model to a single field sector, the
effective inflaton field is boosted and the effective couplings are
suppressed by the number of fields in the model. However, unlike the
simplest models of assisted inflation mostly studied in the
literature, all scalar fields couple to each other. This has important
consequences. First, the effective inflaton field and couplings depend
more strongly on the number of fields in the model, enhancing the
assisted effect. Secondly, the mass spectrum of the ``spectator''
fields, that is the fields orthogonal to the effective inflaton field
that play no role during inflation, is highly non-degenerate.  Both of
these properties have the potential to improve the phenomenology of
the model.

The above is surprising, as studies of assisted inflation have shown
that cross couplings generically kill the assisted behavior. As we
will show in this paper, this conundrum can be solved by noting that
the truncated M-flation model of \cite{Mflation1,Mflation2} is {\it
not} a late-time attractor.  It corresponds to very special initial
conditions. Moreover, despite the increased scaling of the fields and
couplings, the UV behavior of the model is worse than in assisted
models, as higher order operators in the potential grow with the number
of fields in the theory.  Not all is lost though.  Starting with
random initial conditions for all fields, the system generically
evolves towards a late-time attractor solution which is nothing but
quadratic assisted inflation --- assisted inflation with many
uncoupled fields that all have a quadratic potential. 
 
This paper is organized as follows. We will compare M-flation with
assisted inflation, focusing on the role of the spectator fields. The
results for assisted inflation are mostly well known, since they have
been studied extensively \cite{assisted,kanti,Nflation,malik}.  We
repeat them here for easy comparison. In the next two sections we
start with a review of assisted inflation and M-flation respectively,
focusing on the single field truncated sector. In \S \ref{s:UV} we
discuss the UV sensitivity of both models. We find that assisted
inflation does not improve the UV sensitivity with respect to single
field chaotic inflation; the truncated M-flation model, unfortunately,
makes it only worse. Demanding the energy density during inflation to
be below the scale where quantum gravity becomes strong --- lowered
below the Planck scale by the many species in the theory --- bounds
the number of fields to be less than $10^4$.

In \S \ref{s:multifield} we go beyond the truncated sector, and
discuss the full multi-field dynamics (although for a limited number
of fields) in both assisted and M-inflation. In \S \ref{s:multi_ass}
we show that the truncated assisted model is generically a late time
attractor \cite{assisted,kanti}.  \S \ref{s:multi_extended} discusses
M-flation truncated to a 3-field sector. This extension is already
enough to show that the truncated single field model is not an
attractor solution.  Finally, in \S \ref{s:multi_N3} we consider the
evolution of all fields of the matrix model simultaneously (although
for small matrices). The results are in line with what was already
shown in the 3-field model. Provided the system evolves towards the
Minkowski minimum with all fields at the origin, the late-time
attractor is not the truncated M-flation model, but instead quadratic
assisted inflation. In \S \ref{s:conclusions} we end with some
concluding remarks. Appendix \ref{A:perturbations} summarizes the
relevant formulas for the spectrum of density perturbations in
multi-field inflation models. Appendix \ref{A:interactions} lists the
interactions terms for the spectator fields in M-flation.


\section{Assisted inflation}
\label{s:assisted}

In models of assisted inflation many fields evolve simultaneously
during inflation \cite{assisted}. In the simplest set-up the fields do
not couple, and each field experiences the presence of the others only
through the altered background expansion of the universe, which gives
rise to additional damping. Although the potential for each individual
field is too steep to yield inflation, thanks to the enhanced damping a
period of slow roll inflation can nevertheless occur.  In this section
we give a short review of assisted inflation.

Consider a system consisting of $N$ real fields $\phi_i$ with
$i=1,...,N$, with decoupled and equal potentials for each field $V =
\sum_i \tilde V(\phi_i)$ with
\be
\tilde V(\phi_i) = 
\frac{\mu_0^2}{2} \phi_i^2 + \frac{\kappa_0}{3} \phi_i^3 
+ \frac{\lambda_0}{4} \phi_i^4.
\label{Vtilde}
\ee
The Lagrangian is $ \L = \sum_i \( \frac12 \partial_\mu \phi_i
\partial^\mu \phi_i - \tilde V(\phi_i) \)$.  Assuming a spatially
homogeneous and isotropic FRW universe, the equations of motion are
\be
\ddot \phi_i + 3 H \dot \phi_i + \tilde V_{\phi_i}=0,
\quad \forall i
\ee
where $\tilde V_{\phi_i} = \partial_{\phi_i} \tilde V(\phi_i)$.  The Hubble
constant $3H^2 = \sum_i(\dot \phi_i^2/2+ \tilde V)$, which acts as a
damping term in the equations of motion above, scales with the number
of fields $N$.  The larger the number of fields the easier it is to
get slow roll inflation, with all fields slowly rolling down the
potential simultaneously. For equal initial field values, inflation is
effectively driven by a single field, the adiabatic mode.  This can be
made explicit defining
\be
\phi_1 = \phi_0, \qquad
\phi_i = \phi_0 + \psi_i \quad {\rm for} \;i = 2,...,N.
\ee
Truncating to the $(\psi_i =0)$-sector, the action for the
$\phi_0$ mode is
\be
\L = \frac12 N \dot{\phi}_0^2 - \sum_i \tilde V(\phi_0)
\equiv  \frac12 \dot{\phi}^2 - V(\phi)
\ee
where in the 2nd step we introduced the canonically normalized field
$\phi^2 = N \phi_0^2$, and used that $V = \sum_i \tilde V(\phi_0) = N
\tilde V(\phi_0)$. Parameterizing
\be
V =N\(\frac12 \mu_0^2 \phi_0^2 + \frac13 \kappa_0 \phi_0^3 + \frac14 \lambda_0 \phi_0^4\)
= \frac12 \mu^2 \phi^2 + \frac13 \kappa \phi^3 + \frac14 \lambda \phi^4,
\label{V_assist}
\ee
this implies the scaling of the couplings
\be
\mu^2 = \mu_0^2, \quad \kappa = \kappa_0/\sqrt{N}, \quad
\lambda = \lambda_0/N.
\label{scale_ass}
\ee
%

\subsection{Consistent truncation}  
\label{s:trunc_ass}

The truncation to the ($\psi_i =0$)-sector is consistent if this is a
classically stable minimum. The equations of motion for the spectator
fields $\psi_i$ are
\be
\ddot{\psi}_i+ 3H \dot{\psi}_i 
+ \tilde V_{\phi_i}(\phi_0+\psi_i) -  \tilde V_{\phi_i}(\phi_0) =0.
\label{eompsi}
\ee
This is the equation of motion for a scalar field in an effective
potential $V^{\rm eff}_{\psi_i} = \tilde V_{\phi_i}(\phi_0+\psi_i) -
V_{\phi_i}(\phi_0)$ \cite{assisted,malik}. Integrating this expression
gives
\be 
V^{\rm eff} = C + \frac12(\mu^2 + 2 \kappa \phi +3 \lambda \phi^2) 
\psi_i^2 + \frac13 \sqrt{N}(\kappa +3 \lambda \phi)
\psi_i^3 + \frac{N}4 \lambda \psi_i^4 
\label{Veff}
\ee
with $C$ an integration constant. If $V^{\rm eff}_{\psi_i} = 0$ and
$V^{\rm eff}_{\psi_i \psi_i} >0$ at $\psi_i= 0$ this is a classically
stable minimum in which all masses are positive. This is the case for $
(\mu^2 + 2 \kappa \phi + 3 \lambda \phi^2)>0 $.  The effective mass of
the inflaton field and the fluctuations are the same ($V_{\phi\phi} =
V^{\rm eff}_{\psi_i\psi_i}$), hence all fields are light during
inflation. It is expected that the fluctuation fields will move away
from the origin due to quantum fluctuations.

The ($\psi_i=0$)-sector is rather special, as it requires all fields
to have equal initial values. However, it is also a late time
attractor for more general initial values if $\psi_i =0$ is the unique
solution to the equation of motion \eref{eompsi} above
\cite{assisted}.  This is e.g.~the case if the potential is a single
exponent or monomial.  With random initial conditions for all fields,
the slow roll parameters are less than unity and inflation starts as
long as $\sum_i \phi_i^2 \gtrsim \O(1)$.  The multi-field nature of
the model is discussed in more detail in \ref{s:multi_ass}.

\subsection{Inflation in the ($\psi =0$)-sector}

Assisted inflation is truncated to an effective single field inflation
model with a polynomial potential \eref{V_assist}. There are
isocurvature perturbations from the light spectator modes $\psi_i$,
but these do not feed into the curvature perturbation.

The standard chaotic inflation results for a monomial potential $V=
\lambda \phi^p$ are the following. Inflation ends when the slow roll
parameter $\epsilon =1$, which gives $\phi_{\rm end} = p/\sqrt{2} \sim
\O(1)$. See appendix \ref{A:perturbations} for the explicit
definitions of slow roll parameters and the inflationary
observables. Observable scales leave the horizon approximately $\N_* =
60$ e-folds before the end, when $\phi_* = \sqrt{2p\N_*} \sim \O(10)$
--- here, and in the following, the subscript $*$ denotes the
corresponding quantity at horizon exit.  The spectral index is $n_s =1
-(2+p)/(2\N_*)$, and to get the observed power spectrum fixes $\lambda
= \{4 \times 10^{-11},\,2 \times 10^{-12},\, 9 \times 10^{-14}\}$ for
$p=\{2,\,3,\,4\}$. Finally, the ratio of tensor-to-scalar modes is $r
= 4p/\N_*$.  The gravitational wave amplitude is large, and quartic
inflation is marginally excluded as it is $\sim 3\sigma$ away from the
best fit WMAP value; quadratic inflation is still within the $\sim
2\sigma$ range \cite{WMAP}.

Although inflation requires the effective inflaton field to have
super-Planckian field values during inflation $\phi = \O(10)$, the
fields appearing in the original Lagrangian are all below the Planck
scale $\phi_0 < 1$ provided $N > 10^2$ is large enough.  Furthermore,
to get the right amplitude of density perturbations in quartic
inflation requires $\lambda \sim 10^{-14}$ very small, whereas the
original coupling $\lambda_0$ can be larger for large $N$
\eref{scale_ass}.  Note that the suppression of the cubic and
quartic coupling is also beneficial for quadratic inflation, as it may
help explain why the quadratic term dominates during inflation.  It
can however not explain the ratio between the inflaton and Planck mass
$\mu/\mpl \sim 10^{-5}$.


\section{M-flation}
\label{s:mflation}

M-flation is a multi-field model of inflation, in which the inflaton
fields are non-commutative matrices.  In this section we summarize the
results of \cite{Mflation1}, and discuss the M-flation model truncated
to the ``SU(2)-sector'', for which inflation is effectively single
field. M-flation is similar to assisted inflation, in that the fields
and parameters appearing in the effective potential are scaled up/down
by the total number of fields.

The fields in the model are three $N_c \times N_c$ non-commutative
hermitian matrices $\Phi^i$ (we will also use the notation $\vec \Phi
= \{\Phi^1,\Phi^2,\Phi^3\}$), corresponding to $N= 3N_c^2$ real
degrees of freedom. We will refer to $N_c$ as the number of colors.
The Lagrangian is $\L = \L_{\rm kin} -V$ with
\ba
\L_{\rm kin} &=& 
\frac12 \Tr \( D_\mu \Phi^i D^\mu \Phi_i \)
\nn \\
V &=& \Tr \(
\frac{\mu_0^2}{2} \Phi^i \Phi^i 
-\frac{i\k_0}{6} \eps_{ijk} [\Phi^i,\Phi^j]\Phi^k
-\frac{\lambda_0}{8} [\Phi^i,\Phi^j][\Phi^i,\Phi^j]\)
\label{L1}
\ea
with repeated indices $i,j =1,2,3$ summed over.  The Lagrangian is
invariant under a global $SU(2)$ symmetry acting on the indices $i,j$,
and a global or local $SU(N_c)$ symmetry acting on the matrices.  In
the latter case the covariant derivative contains a gauge
connection. The above Lagrangian is motivated from string theory, it
arises as the world-volume theory of $N_c$ coincident $D3$-branes
\cite{Mflation1}.

We decompose $\vec \Phi$ into a scalar ``trace field'' $\phi_0$ and
$3N_c^2-1$ ``spectator fields'' contained in the three matrices $\vec
\Psi$ via \cite{Mflation1}
\be
\vec \Phi = \phi_0 \vec J + \vec \Psi,
\label{phi}
\ee
with $J^i$ are the generators of the $N_c$-dimensional irrep of
SU(2) which satisfy the usual relations
\be
[J_i,J_j] = i \eps_{ijk}J_k, \qquad
\Tr(J_i J_j) = N_c(N_c^2-1)/12\delta_{ij}
\equiv \frac13 d_c \delta_{ij}  .
\ee
It follows that $\phi_0 = 1/(d_c) \Tr(\Phi^i J^i)$ and $\Tr(\Psi^i
J^i)=0$; the trace field is aligned with $\vec J$, whereas the
spectator fields are orthogonal to it. With this decomposition the
kinetic term becomes
\be 
\L_{\rm kin} = \frac12 d_c (D_\mu \phi_0)^2 + \frac12
\Tr(D_\mu \vec \Psi)^2 .
\label{Lkin1}
\ee
The effectively single-field truncated SU(2)-sector corresponds to
$\vec \Psi =0$; in this limit the potential reads
\be
V^{(0)} = d_c\(
 \frac{\mu_0^2}{2}  \phi_0^2
+\frac{\k_0}{3} \phi_0^3
+ \frac{\lambda_0}{4}  \phi_0^4\)
=
\( \frac{ \mu^2}{2}  \phi^2
+\frac{\k}{3}  \phi^3
+ \frac{ \lambda}{4}  \phi^4\).
\label{V0}
\ee
In the 2nd step we introduced the canonically normalized field $\phi$
via
\be
\phi^2 = d_c \phi_0^2
\label{rescale}
\ee
and factored out factors of $d_c$ in the definition of the couplings:
\be
\mu^2 = \mu_0^2,\quad
\k = \frac{\k_0}{\sqrt{d_c}}, \quad
\lambda = \frac{\lambda_0}{d_c}.
\label{rescale2}
\ee
In the large $N_c$ limit $d_c \propto N_c^3 \propto N^{3/2}$ with
$N_c$ the number of colors and $N$ the number of fields.  Comparing
with assisted inflation \eref{scale_ass}, we see that the effective
inflaton field and quartic coupling scale with $N^{3/2}$ instead of
$N$. The assistance of the spectator fields is more efficient. For
example, to have $\lambda_0 \sim 1$ in quartic chaotic M-flation
requires $N \sim 10^9$, whereas it would require $N \sim 10^{13}$ in
assisted inflation.

\subsection{Consistent truncation}

The truncation to the $SU(2)$-sector is consistent provided the
potential contains no terms linear in $\vec \Psi$, and all the modes
have a positive definite mass.  In that case, starting with $\vec \Psi
= \dot {\vec \Psi} =0$, classically the spectator fields remain at the
origin.

The linear potential is
\be
V^{(1)} = \frac1{\sqrt{d_C}}\[2 \lambda \phi^3 
+\frac{2 \kappa}{3} \phi^2 + \frac{\mu^2}{2} \phi
\]{\rm Tr}(\Psi^j J^j) =0,
\ee
which indeed vanishes as $\vec \Psi$ has no components along $\vec J$,
and the above trace is zero.  To determine the mass spectrum of the
spectator fields one needs to diagonalize the quadratic potential.
This can be done introducing eigenvectors $\vec \Psi = \sum_w \vec
\psi_w$ which satisfy
\be
i\eps_{ijk} [J^i,\psi_w^j] = w \psi_w^k.
\label{ev}
\ee
The quadratic potential becomes
\be
V^{(2)} = \frac12 
\( \frac{\lambda}{2} \phi^2 w (w-1) 
- \k \phi w + {\mu^2} \) {\rm Tr}(\psi_w^i \psi_w^i)
\equiv \frac12 \mu_w^2{\rm Tr}(\psi_w^i \psi_w^i)
\label{V2}
\ee
The mass eigenstates $\psi_w$ are solutions of the eigenvalue equation
\eref{ev} for a specific eigenvalue $w$. They can be further expanded
in spherical harmonics of $SU(2)$. Since the action of $J^i$ on the
spherical harmonics is known, the eigenvalue equation \eref{ev} can be
solved explicitly.  The details can be found in
\cite{Mflation1,raamsdonk}; here we just summarize the results.  The
solutions come in three classes: $w = -1, -(l + 1), l$, corresponding to
so-called zero, $\alpha$ and $\beta$ modes respectively.  We write
$\{\iota, \alpha, \beta\}^i_{lm}$, with as before, $i=1,2,3$ the SU(2)
index, and $l,m$ integer valued angular momenta indices.

\begin{enumerate}
\item Zero modes $\vec \iota_{lm}$ have $w = -1$, while $1\leq l \leq
N_c-1$, and $-l \leq m \leq l$.  From \eref{V2} the zero mode mass is
$\mu_\iota^2 = V_\phi^{(0)}/{\phi}$, which vanishes in a $SU(N_c)$
breaking minimum with $\phi \neq 0$; the zero modes are the
corresponding Goldstone bosons.  If the $SU(N_c)$ symmetry is gauged,
the Goldstone bosons are eaten by the gauge fields, and there are
$N_c^2-1$ massive vector bosons in the spectrum with mass $m_A \sim g
\phi$ and $g$ the gauge coupling. Note that during inflation the
$SU(N_c)$ symmetry is broken spontaneously.

\item
$\alpha$-modes $\vec \alpha_{lm}$ have $w = - (l+1)$, $ 1 \leq l \leq
N_c-1$ and $-(l-1) \leq m \leq (l-1)$. Each $l$-mode has a
$(2l-1)$-fold degeneracy, for a total of $1-2N_c+N_c^2$ modes.
However, the mode $\vec \alpha_{10} \propto \vec J$ (see \eref{modes}
for the explicit definition) is nothing but the trace field $\phi$.
The mass of the ($l=1$)-state equals the $\phi$-mass, as it should for
this identification to work. Thus the number of $\alpha$ modes is
$N_c(N_c-2)$.

\item
$\beta$-modes $\vec \beta_{lm}$ have $w = l$, $0 \leq l \leq N_c-1$
and $-(l+1) \leq m \leq (l+1)$.  Each $l$-mode has a $(2l+3)$-fold
degeneracy, for a total of $N_c(N_c+2)$ modes.
\end{enumerate}
For a global $SU(N_c)$ system the total number of spectator modes
$\vec{\psi}_w$ is $(N_c^2-1) + N_c(N_c-2)+ N_c(N_c+2) =
3N_c^2-1$.  Together with the trace field $\phi$ this makes up the
$N=3N_c^2$ d.o.f.~of three $N_c \times N_c$ hermitian matrices $\vec \Phi$.
For a local $SU(N_c)$ system the zero modes are eaten during
inflation, and there are $N=2N_c^2+1$ scalar d.o.f. corresponding to
the $\alpha$, $\beta$ modes and the trace field.  In addition there
are $N_c^2-1$ massive gauge fields with $m_A \sim g \phi$.

If all masses are positive definite during inflation, it is
classically consistent to set $\vec \psi_w =0$.  For heavy fields
$\mu_w^2 \gtrsim H^2 \approx V_0/3$ the origin is stable quantum
mechanically as well.  However, light fields will quantum fluctuate
during inflation, moving away from the origin by a random walk
effect. The zero modes have a mass comparable to the inflaton mass,
and thus they are all light.  The $\alpha$ and $\beta$ mode spectrum
is non-degenerate; only the states with small angular momentum are
light.

\subsection{Trace inflation}
\label{s:traceflation}

The multi-field matrix model reduced to the SU(2)-sector yields single
field inflation. In a string theory set-up the action arises from a
stack of $N_c$ D3-branes in a curved background; in these models the
$\kappa$-parameter is not independent \cite{Mflation1,Mflation2,raamsdonk}:
\be
\kappa = -3  \sqrt{\lambda/2} \mu.
\label{kappa}
\ee
This brings the potential in the ``symmetry breaking'' form
\be
V = \frac14 \lambda \phi^2(\phi-\bar \mu)^2,
\qquad {\rm with} \; \bar \mu = \mu\sqrt{2/\lambda}.
\label{Vsymmb}
\ee
For definiteness, in the remainder of this paper we will concentrate
on this specific potential; the generalization to generic
$\kappa$-values is straightforward. We will call this model trace
inflation, as it only involves the trace field.  The potential
\eref{Vsymmb} has two minima at $\phi_{\rm min} =0,\bar \mu$, with a
maximum in between at $\phi_{\rm max} = \bar \mu/2$. It is symmetric
under $\phi \to -\phi + \bar \mu$, that is symmetric around $\phi_{\rm
max}$.

Taking $\bar \mu \ll \phi_*$, the $\bar \mu$-term can be neglected
during observable inflation and the potential is approximately quartic
$V = (\lambda/4) \phi^4$. This gives the usual result for the spectral
index and tensor ratio.  In the opposite limit $\bar \mu \to \infty$
(but $\lambda \bar \mu^2$ kept finite) the potential around both
minima is approximately quadratic with corrections suppressed by
factors $\phi_*/\bar \mu$.  We can expand the potential during
inflation around the origin in small $\phi_*/\bar \mu$ 
\footnote{The expansion around the $\phi_{\rm min} = \bar \mu$ minimum
gives the same result. The corresponding slow roll parameters can be
found applying the symmetry $\phi \to -\phi + \bar \mu$ to
\eref{sr_corrections}.}; the result for the slow roll parameters is
\be
\eps_* = \frac{2}{3\phi_*^2}\(1-2 \frac{\phi_*}{\bar \mu}+ 
 ...\),
\quad
\eta _*= \frac{2}{3\phi_*^2}\(1-4 \frac{\phi_*}{\bar \mu}+ 
...\)
\label{sr_corrections}
\ee
with the ellipses denoting higher order corrections in $\phi_*/\bar
\mu$. The first order corrections cancel in the expression for the
spectral index $n_s = 2\eta - 6\eps$, which gets only corrected at
second order.  However, the tensor to scalar ratio $r = 16 \eps$ is
already affected at first order.  While the deviations from a
quadratic potential are small for the spectral index, less than
$0.1\%$, the corrections to $r$ are appreciable, of order $10\%$ for
$\bar \mu = 10 -10^3$, and the potential is distinguishable from a
purely quadratic potential.  Large deviations from either a quartic or
quadratic potential are only expected for intermediate values $\bar
\mu \sim \phi_* = \O(10)$.

Inflation occurs for super-Planckian field values $\phi \sim
\O(10)$. This corresponds to sub-Planckian values for the field
$\phi_0$ appearing in the original Lagrangian provided $N_c >10$. As
mentioned before, to have order one couplings $\lambda_0, \kappa_0$
requires a large number of fields $N \sim N_c^2 \sim 10^9$.

Just as in assisted inflation there is no fine-tuning of the initial
conditions needed to get inflation.  As long as $\phi^2 + \sum |\vec
\psi_w|^2 \gg 1$, the slow roll parameters are small, and inflation
will start. Because of all the cross couplings between the fields in
the potential, it is not straightforward to determine whether the
SU(2)-sector is a late time attractor solution, and thus a sufficient
description to describe the last 60 e-folds of inflation. We return to
these issues in \S \ref{s:multifield}.

\section{Number of spectator fields}
\label{s:UV}

Both perturbative \cite{veneziano,arkani} and non-perturbative
\cite{dvali1,dvali2} arguments suggest that the scale where quantum
gravity becomes strong is lowered in theories with many
particles. This has implications for the UV behavior of assisted
inflation and M-flation \cite{Nflation}. Even though in these models
all scalar fields have sub-Planckian field values during inflation,
the amplitudes may nevertheless exceed the effective quantum gravity
scale. In addition bounds on the number of field in thesed models can
be derived.

We restore explicit factors of $\mpl$ in this section.

The presence of a large number of fields makes gravity parametrically
weaker. Radiative stability of Newton's constant suggest that the
scale where quantum gravity becomes strong is lowered to
\cite{veneziano,arkani}
\be
\Lambda^2 = \frac{\mpl^2}{N_{\rm cl}},
\label{cutoff}
\ee
where $N_{\rm cl}$ counts all the species with mass below the cutoff
$\Lambda$.  This may for example be deduced from one-loop corrections
to the graviton propagator. The quadratic divergence can be absorbed
in a redefinition of the Planck mass, and barring accidental
cancellations this yields $ \mpl^2 \sim N_{\rm cl} \Lambda^2$ with
$N_{\rm cl}$ the number species running in the loop.  The same cutoff
follows from non-perturbative considerations of classical black holes
\cite{dvali1,dvali2}. In particular, black holes of size
$\Lambda^{-1}$ with the cutoff given in \eref{cutoff} are no longer
semi-classical, as they have lifetime $\tau_{\rm BH} \sim
\Lambda^{-1}$ shorter than their size.  Species with a mass exceeding
the cutoff $\Lambda$ cannot be treated semi-classically, and do not
enter the arguments.

\subsection{Bounds on the number of species}

A strong, and model-independent bound, on the number of species can be
derived requiring that the energy density driving inflation is below
the cutoff scale \eref{cutoff}.  For chaotic inflation $V_* \sim H_*^2
\mpl^2 \sim 10^{-8} \mpl^4$, and thus
\be
V_*^4 < \Lambda^4
\quad \Leftrightarrow \quad
N_{\rm cl} < 10^4
\label{boundN}
\ee
For larger $N_{\rm cl}$ the low energy effective description breaks
down and a UV completion is needed.  With this bound satisfied, all
$\alpha$ and $\beta$ modes of M-flation have mass below the cutoff,
and $N_{\rm cl} \sim N$. Indeed, the heaviest state $\vec \psi_w$ has
$w \sim N_c$ with $\mu_w^2 \sim \lambda \phi^2 N$ \eref{V2}, and thus
\be
\mu_{(w \sim N_c)}  < \Lambda 
\quad \Leftrightarrow \quad
N_{\rm cl} < 10^5 f.
\label{maxmass}
\ee
Here $f = 10^{-5}/(\sqrt{\lambda} \phi) \geq 1$ which is saturated if
the potential during inflation is dominated by the quartic term.  If
\eref{boundN} is satisfied, \eref{maxmass} is automatically satisfied
as well.  All $\alpha$ and $\beta$ modes, and in the ungauged model
also the zero modes, have mass below the cutoff $\Lambda$.  In gauged
M-flation the zero modes are eaten by the gauge bosons which pick up a
mass $\mu_A \sim g \phi_* \gtrsim \mpl$ during inflation (and possibly
$\mu_A \sim g \bar \mu$ after inflation if the inflaton settles in the
$\phi = \bar \mu$ minimum).  The heavy gauge fields may be integrated
out consistently, and the low energy effective theory is just the
ungauged model but without the zero-modes.

Other bounds on the number of species can be found in the literature
\cite{lim,ahmad,huang}, but these are all much weaker. For example,
the gradient energy of the light quantum fluctuating spectator fields
is bounded to be less than the potential energy driving inflation
\cite{gradient,dvalilust}:
\be 
\sum (\nabla \vec \psi_w)^2 < V_* 
\quad \Leftrightarrow \quad
N_{\rm light} < 8\pi^2 \( \frac{\mpl^2}{H_*^2} \) \sim 10^{10}
\label{maxE}
\ee
with $N_{\rm light}$ the number of light fields with $\mu_w^2 < H^2$.
In assisted inflation all fields and in ungauged M-flation all zero
modes are light, and $N_{\rm light} = N$. However, in gauged M-flation
only the lowest lying $\alpha$ and $\beta$ modes are light, and
\eref{maxE} is easily satisfied. Bounds of similar magnitude can be
derived from black hole arguments.  The cutoff \eref{cutoff} can be
rewritten as a bound on the number of species, by noting that all
particles $N_{\rm cl}$ have mass $\mu_w < \Lambda$.  For a degenerate
mass spectrum this gives \cite{dvali1}
\be N  < (\mpl/\mu)^2
\label{Ncl}
\ee
Applying this bound to the post-inflationary vacuum, even the most
constrained gives only $N < 10^{10}$.

\subsection{UV behavior}

As in every single field chaotic inflation model, the effective
inflaton field in assisted inflation and M-flation has super-Planckian
values during inflation. However, the fields appearing in the original
Lagrangian all have values below the Plank scale.  This has led to
assertions that corrections from higher order operators are small, and
the UV behavior of the theory is under control \cite{assUV}.  As we
have seen in the previous subsection, in theories with many species,
the effective cutoff where quantum gravity effects become strong is
lowered \eref{cutoff}.  This suggest that for a good UV behavior the
requirement is that all field amplitudes, or at least the energy scale
during inflation $V_*^{1/4}$, is below the cutoff $\Lambda =
\mpl/\sqrt{N_{\rm cl}}$ rather than below the Planck scale
\cite{Nflation,arkani,linde}.  In this subsection we discuss the UV
behavior of assisted inflation and M-flation, starting with the
former.

\paragraph{Assisted inflation}
So far we have only considered the renormalizable part of the
potential. Working in an effective field theory set-up, one needs to
include all non-renormalizable operators consistent with the
symmetries of the model.  These additional operators are suppressed by
some cutoff $\Lambda$, which is the scale of new physics. The full
potential for the field $\phi_i$ is
\be
\tilde V(\phi_i) \supset 
\alpha_d \frac{\phi_i^{4+d}}{\Lambda^d} .
\ee
For the effective single field potential $V = N\tilde V$ this
translates to
\be
V = \frac12 \mu_0^2 \phi^2 + \frac{\kappa_0}{\sqrt{N}}
\phi^3 + \frac{\lambda_0}{N} \phi^4 + 
\sum_{d \geq 1} \frac{\alpha_d}{N} \phi^4
\(\frac{\phi^2}{\Lambda^2 N}\)^{d/2}.
\ee
Assuming $\alpha_d \sim \O(1)$, higher order operators are small
provided
\be \frac{\phi^2}{N\Lambda^2} \ll 1.
\label{nonren}
\ee
During chaotic inflation $\phi = \O(10)\mpl$. If one takes the cutoff
at the Planck scale $\Lambda = \mpl$, then the whole series of
non-perturbative operators is under control in the large $N \gg 10^2$
limit \cite{assUV}. However, as discussed, in a theory with many
particles the scale where quantum gravity effects become large is
lowered \cite{Nflation}. Taking \eref{cutoff} as the cutoff,
\eref{nonren} exceeds one, and {\it all} higher order terms are large
during inflation. The theory is extremely UV sensitive, and control
over higher order operators is lost. In this respect assisted
inflation does not improve over non-assisted single field chaotic
inflation, as in both set-ups the inflaton field exceeds the effective
cutoff during inflation ($\phi_0 > \Lambda =\mpl/\sqrt{N}$ in assisted
inflation, and $\phi > \mpl$ in single field chaotic inflation
respectively).

Above we assumed that the coefficients of the higher order terms are
$\alpha_d \sim \O(1)$.  Although this seems a reasonable assumption,
it is not necessarily true.  Linde argues \cite{linde} that the
constant part of the scalar field does not appear in the gravitational
diagrams directly, but only via its effective potential $V$ and the
masses of particles interacting with the scalar field $\phi$.  Thus
the expansion should be in $V/\Lambda^4$ or $m(\phi)^2/\Lambda^2$
rather than in $\phi/\Lambda$.  An explicit example for which this
reasoning applies is (the low energy limit of) a supergravity model of
chaotic inflation, where the inflaton has a shift symmetry
\cite{kawasaki}. If true, the UV behavior is under control as long as
the number of fields is bounded \eref{boundN}.

It follows that for either an expansion in $\phi/\Lambda$ or in
$V/\Lambda^4$ the control over UV physics is the same in single-field
and in multi-field assisted chaotic inflation. With this in mind the
main (and only) advantage of assisted inflation is that the couplings
are suppressed by factors of $N$, and do not have to be as small as in
standard chaotic inflation.  However, the assisted behavior is limited
\eref{boundN}; the quartic coupling $\lambda = \lambda_0/N$
\eref{scale_ass} can be suppressed by at most a factor $10^4$.

\paragraph{M-flation}

To asses the UV behavior of M-flation, write down all renormalizable
and non-renormalizable operators consistent with all symmetries:
\be
V =
\sum_{d=1}
{\rm Tr}\(
\frac{\alpha_{d}}{\Lambda^{2(d-1)}} (\Phi^i \Phi^i)^{d+1}
+\frac{\kappa_d}{\Lambda^{3d-1}} (\eps_{ijk} [\Phi^i,\Phi^j]\Phi^k)^{d+1}
+\frac{\lambda_d}{\Lambda^{4d}}([\Phi^i,\Phi^j][\Phi^i,\Phi^j])^{d+1}
+... \)
\ee
Suppressing order one coefficients, and writing the potential in terms
of the canonical normalized field $\phi$ gives
\ba V &=& 
\alpha_0 \Lambda^2 \phi^2 + \frac{\kappa_0}{\sqrt{d_c}} \phi^3 +
\frac{\lambda_0}{d_c} \phi^4 
\nn \\&+& \sum_{d=1} \(
\alpha_{d} \Lambda^2 \phi^2
\(\frac{\phi^2}{\Lambda^2} \)^{d}
+ \frac{\kappa_d \Lambda}{\sqrt{d_c}} \phi^3
\(\frac{\phi^3}{\sqrt{d_c}\Lambda^3} \)^{d}
+ \frac{\lambda_d}{d_c} \phi^4
\(\frac{\phi^4}{d_c \Lambda^4} \)^{d}
+...\)
\ea 
where we can identify $\mu_0^2 =\alpha_0 \Lambda^2$. The ellipses
denote products of the $\alpha,\kappa$ and $\lambda$-terms.  For now,
assume $\alpha_d,\kappa_d,\lambda_d \sim \O(1)$.

First take the cutoff at the Planck scale $\Lambda = \mpl$.  All
$\kappa_d$ and $\lambda_d$ interactions are under control during
inflation if $d_c > (\phi/\mpl)^6 \sim 10^6$ or $N_c > 10^2$; the
strongest constraint comes from the $\kappa_d$-terms. But the
$\alpha_d$ terms are problematic, as they are not suppressed by
factors of $N_c$.  Hence, unless there is a UV completion in which
these terms are absent, or unnaturally small, they ruin inflation.  In
fact, all higher order operators containing factors of $(\Phi_i
\Phi_i)^d$ have the same problem. This problem was already apparent in
the renormalizable inflaton potential.  Although $\kappa_0$ and
$\lambda_0$ can be large, as the effective couplings \eref{rescale2}
are suppressed by powers of $N_c$, this is not the case for the mass
term $\mu = \mu_0$.  To get inflation one has to tune the hierarchy
$\mu \lesssim 10^{-5} \mpl$.

The situation gets worse if one takes into account that the effective
cutoff where quantum gravity effects become important is lowered below
the Planck scale by the many particles in the theory \eref{cutoff}. In
this case the $\kappa_d$-terms are proportional to
$(\sqrt{N_c}\phi/\mpl)^{3d}$, while the $\lambda_d$ terms $\propto
(N_c^{1/4}\phi/\mpl)^{4d}$.  The UV behavior is even worse than in
single field chaotic inflation, as higher order terms grow with
powers of $N_c$ in addition to the usual powers of $(\phi/\mpl)$.

Once again, if following \cite{linde} the expansion should be in
$V/\Lambda^4$ rather than $\phi/\Lambda$ the UV behavior is under
control for $N < 10^4$ \eref{boundN}.  

To conclude, just as for assisted inflation, the main and only
advantage of M-flation over single-field chaotic inflation is that the
couplings are suppressed by the number of fields. In M-flation the
quartic coupling $\lambda = \lambda_0/N^{3/2}$ \eref{rescale2} can be
suppressed by at most a factor $10^6$. But there is no gain in control
over the non-renormalizable operators; on the contrary, M-flation may
make the tuning only worse.


\section{Multi-field inflation}
\label{s:multifield}

So far we have focused on a truncated sector of assisted inflation and
M-flation, in which inflation is effectively single field. If the
truncated sector is an attractor solution the single field description
is a good approximation for the last 60 e-folds of inflation (provided
the total number of e-folds is much larger than 60), when observable
scales leave the horizon.

If inflation in the truncated sector is {\it not} an attractor
solution, the evolution of all fields should be taken into account for
a correct description.  The extra fields can change the dynamics with
respect to truncated inflation in two ways. First, isocurvature
perturbations may feed into the curvature perturbation, which yields
potentially measurable deviations from the single field consistency
relation \eref{sinD}. Multifield dynamics can also enhance
non-Gaussianity. For assisted inflation and M-flation non-Gaussianity
is suppressed by the smallness of the slow roll parameters and is
negligible. Second, even if observable inflation is effectively single
field and the isocurvature modes can be neglected (and unfortunately
the multi-field dynamics cannot be probed directly), the effective
potential is generically different than that of the truncated
sector. The predictions for the inflationary observables such as the
spectral index and the tensor-to-scalar ratio will differ.  What is
more, the nature of the assisted behavior, and the scaling of
couplings and field with the number of fields, may differ.

The details concerning perturbations in multi-field inflation, and the
inflationary observables, can be found in appendix
\ref{A:perturbations}.

\subsection{Assisted inflation}
\label{s:multi_ass}

\begin{figure}[ht]
\centering \subfigure[Assisted inflation with $\phi_{\rm max} = 10$,
$\bar \mu =200$, giving $n_s = 0.967$.]
{\includegraphics[scale=.65]{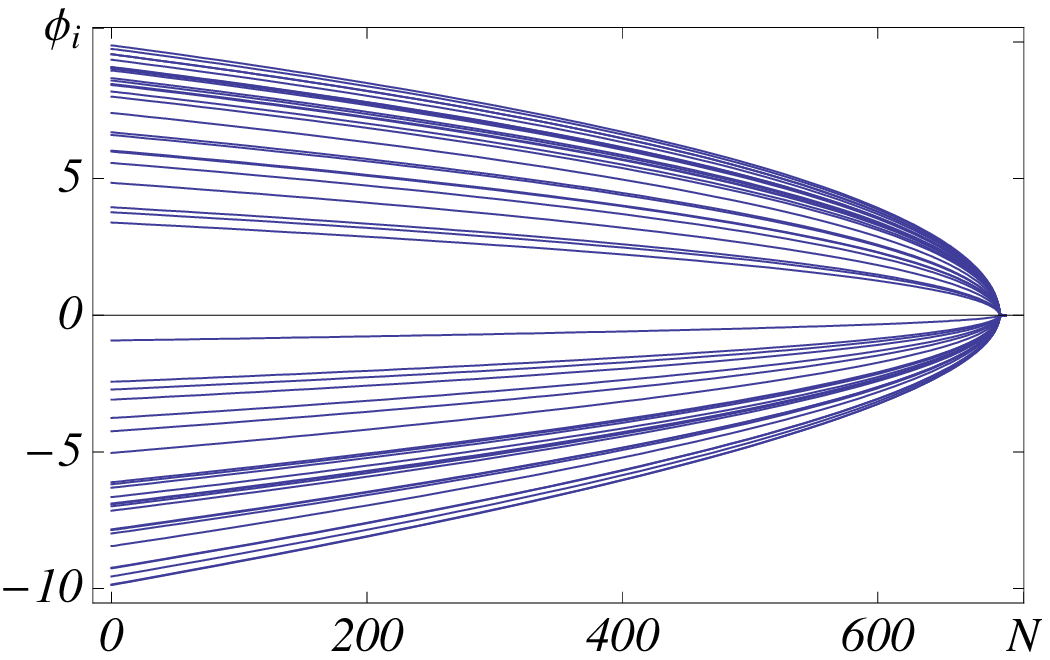}
\label{F:ass1}}\hspace*{0.5cm}
\subfigure[M-flation with $\phi_{\rm max} = 10^2$,
$\bar \mu =1$, giving $n_s = 0.967$.]
{ \includegraphics[scale=.65]{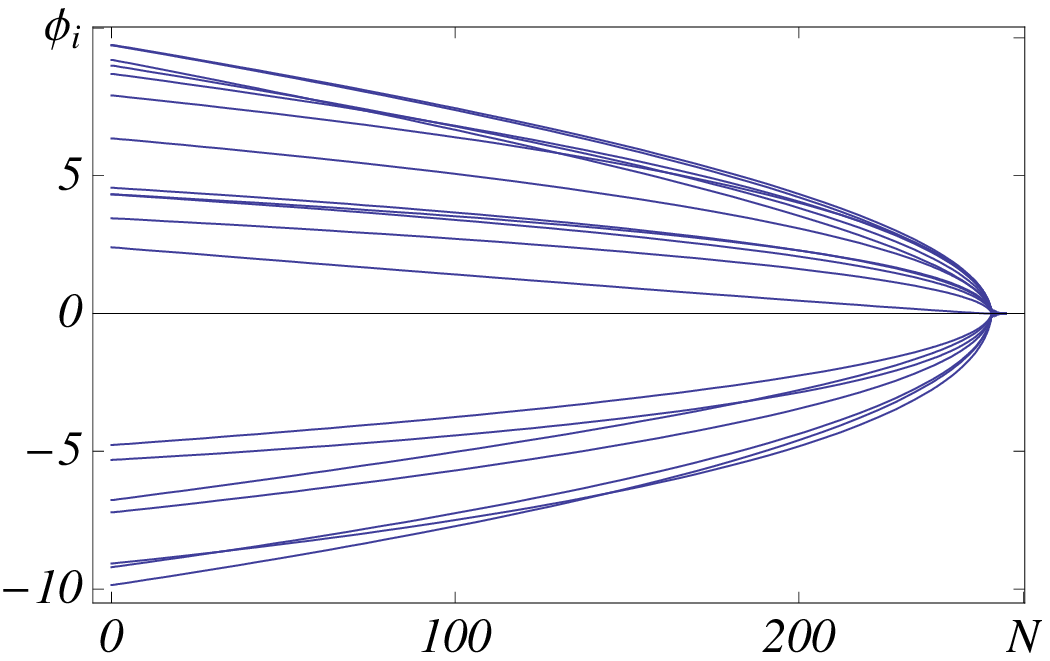}
\label{F:M3}}
\newline
\centering \subfigure[Assisted inflation with $\phi_{\rm max} = 10^2$,
$\bar \mu =1$, giving $n_s = 0.95$. ]
{\includegraphics[scale=.65]{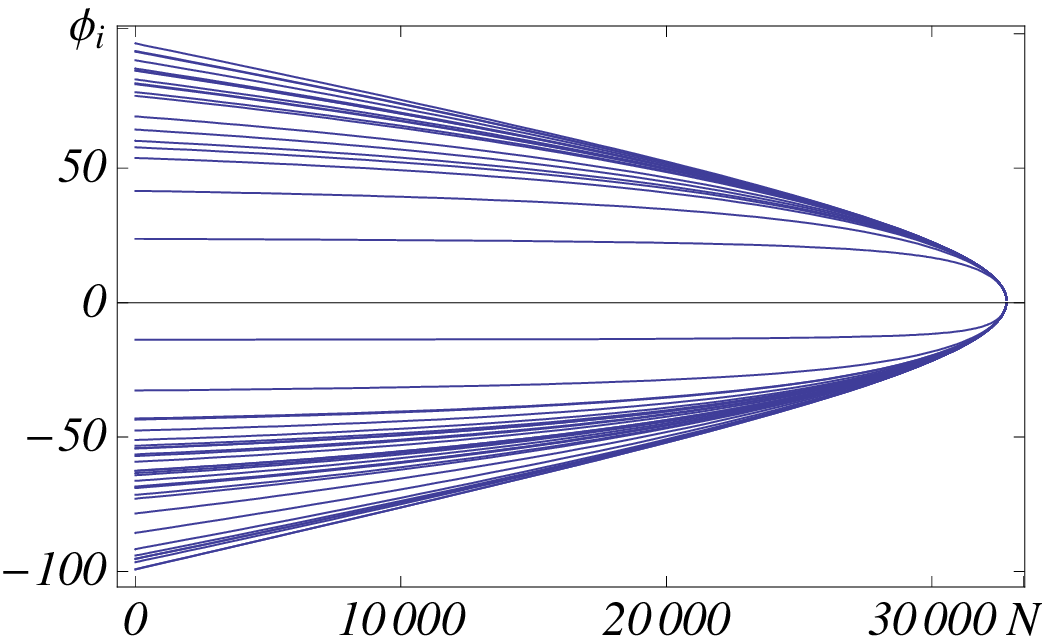}
\label{F:ass2}} \hspace*{0.5cm}
\subfigure[M-flation with $\phi_{\rm max} = 10^2$,
$\bar \mu =1$, giving $n_s = 0.967$.]
{ \includegraphics[scale=.65]{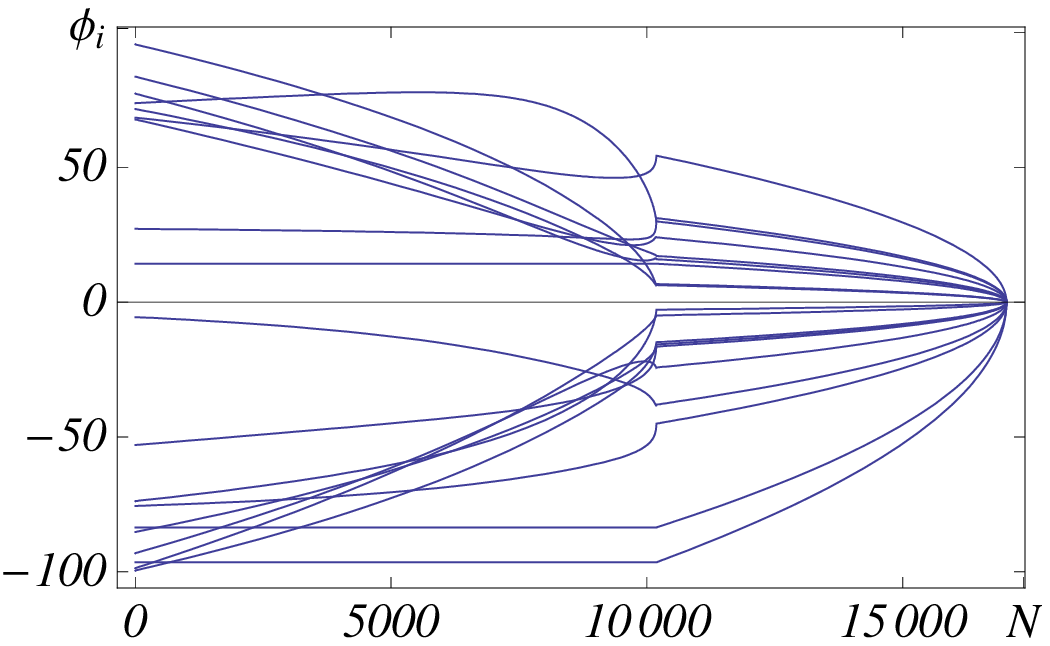}
\label{F:M2}}
\newline
\centering
\subfigure[Assisted inflation with $\phi_{\rm max} = 10^2$,
$\bar \mu =40$.]
{\includegraphics[scale=.65]{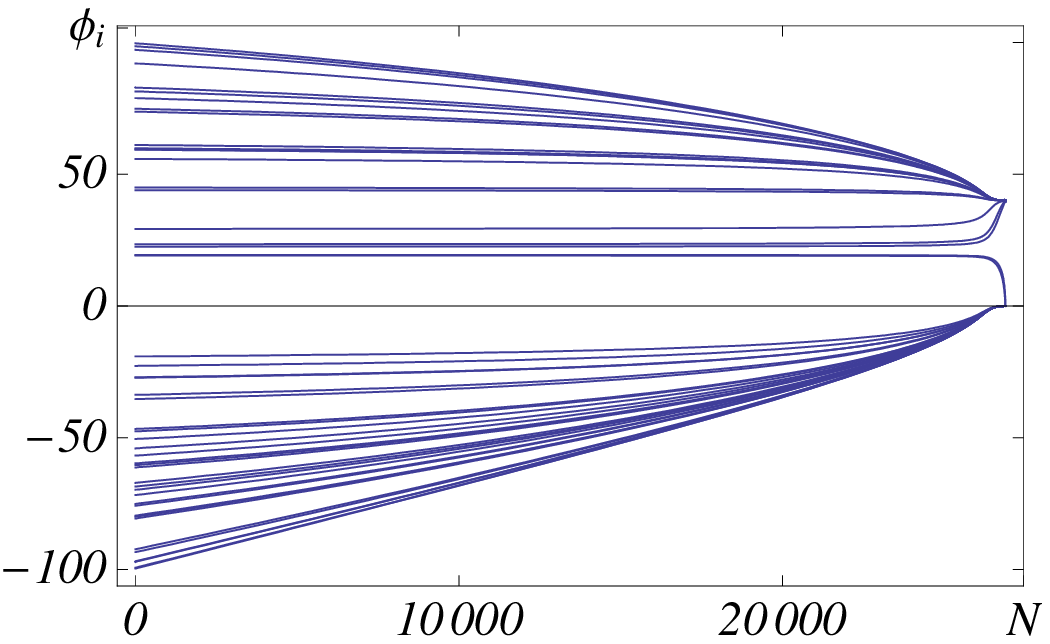}
\label{F:ass3}} \hspace*{0.5cm}
\subfigure[M-flation with $\phi_{\rm max} = 60$, $\bar \mu =40$,
evolving the wrong minimum with $V<0$.]  {
\includegraphics[scale=.65]{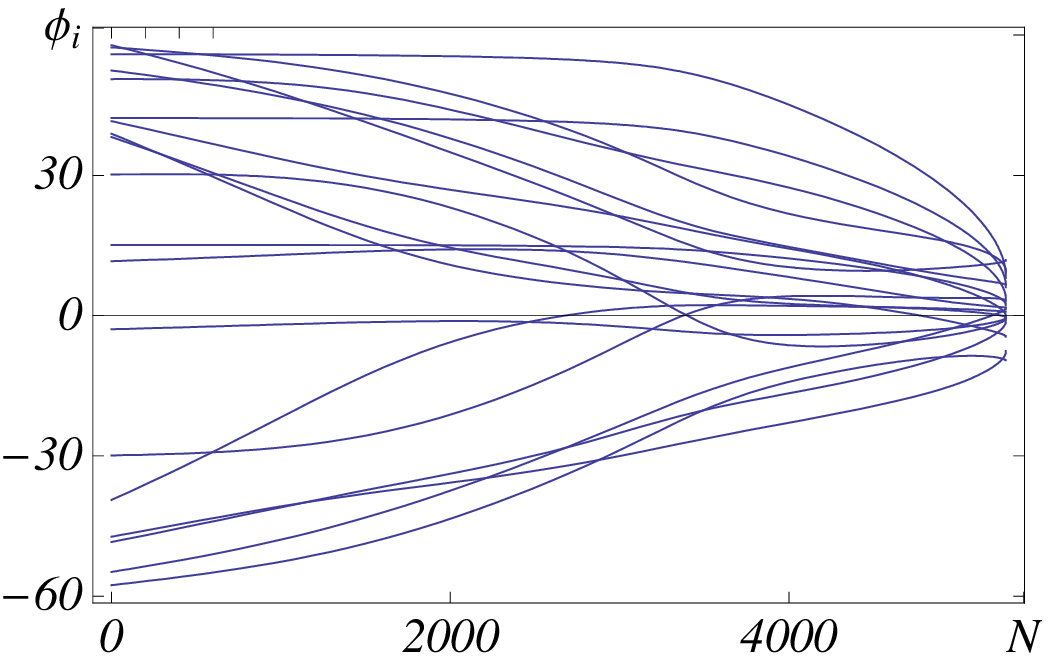}
\label{F:M1}}
\label{F:assM}
\caption[]{Evolution of fields in assisted inflation ($N=50$) and
M-flation ($N_c=3$, gauged, $N=19$) respectively, as a function of the
number of e-folds $\N$. Initial conditons are choosen from a flat
distribution for $(\phi_i)_\ini^2$ with $(\phi_i)_\ini \in [-\phi_{\rm
max}, \phi_{\rm max}]$, see footnote \ref{distribution}.}
\end{figure}

As discussed in \S \ref{s:trunc_ass} assisted inflation can be
consistently truncated to the ($\psi =0$)-sector, corresponding to the
set-up where all fields have equal initial amplitudes. As is well
known, this sector is a late-time attractor if $\psi_i=0$ is the
unique solution to the equations of motion for the spectator fields
\eref{eompsi}, which happens for a simple monomial or exponential
potential \cite{assisted}.

For a quadratic potential it is clear that inflation is effectively
single-field, as one can always make a rotation in field space, such
that the effective inflaton field is the only (canonically normalized)
field with non-zero field value. Fig.~\ref{F:ass1} shows the evolution
as a function of number of e-folds $\N$ for $N=50$ fields in quadratic
assisted inflation. All fields are uncoupled, and have a degenerate
mass; consequently their relative amplitudes remain constant with
time. For all plots in Fig.~1 we choose random initial conditions for
all fields from the interval $(\phi_i)_\ini \in [-\phi_{\rm
max},\phi_{\rm max}]$, with a flat distribution for $(\phi_i)_\ini^2$
with all values equally likely~\footnote{\label{distribution} The
probability density function for the initial amplitudes is taken
$f(\phi_\ini) = \phi_\ini/\phi_{\rm max}^2$ with support $\phi_\ini
\in [-\phi_{\rm max},\phi_{\rm max}]$.  The corresponding probability
distribution for $\phi_\ini^2$ is flat: $\tilde f(\phi_\ini^2) =
1/\phi_{\rm max}^2$ with support $\phi_\ini^2 \in [0,\phi_{\rm
max}^2]$.} .

In quartic inflation, that is set $\mu_0,\kappa_0 =0$ in
\eref{Vtilde}, the field with largest amplitude starts rolling
first. As soon as it falls below the field with the next-to-largest
amplitude, this field starts rolling as well, and both continue in
unison.  This process continues, and during the last 60 e-folds all
fields with an initially large amplitude roll together, a situation
very well approximated by the truncated model. All fields with a small
amplitude have not started rolling yet, and effectively decouple from
assisted inflation.\footnote{The effective inflaton is $\phi =
\sqrt{N} \psi$ with $N$ the number of rolling fields, and $\psi$ the
(equal) amplitude of these individual fields.  Fields with amplitude
$\psi_i \lesssim \phi_{\rm end}/\sqrt{N}$ remain frozen during
inflation.}  If inflation lasts much longer than 60 e-folds and the
number of fields is large, the attractor solution will be reached
before the end of inflation, and the truncated description applies.
In the opposite limit, measurable deviations from single field
inflation are possible.  The reason is that the number of fields that
only start rolling during the last 60 e-folds, and which act as
perturbations on the adiabatic mode corresponding to all fields that
started rolling before, is relatively large. Fig.~\ref{F:ass2} shows
the evolution of $N=50$ fields in quartic assisted inflation. The
assisted behavior can be seen.

For the symmetry breaking potential \eref{Vsymmb} $\psi_i=0$ is not
always an attractor. Large deviations may occur for $\bar \mu \gtrsim
\phi_* = \O(10)$, but smaller than the majority of the initial
amplitudes of the fields.  In this case, fields with initial
amplitudes $|\psi_i|_{\rm in} > \bar \mu$ or $(\psi_i)_{\rm in} < 0$
start rolling first, and may settle in their respective minima before
the end of inflation.  The fields with amplitude $0 \lesssim
(\psi_i)_{\rm in} \lesssim \bar \mu$, if present, are the only ones
rolling during the last 60 e-folds.  If many, their motion can still
be described by a truncated assisted model, but now the number of
fields $N$ is greatly reduced, and the initial amplitude of the
participating fields is bounded.  Clearly, simply truncating the {\it
full} system would give the wrong results.  Fig.~\ref{F:ass3} shows the
evolution of the fields in assisted inflation with a symmetry breaking
potential. Fields with a large amplitude settle in their minimum
before the last 60 e-folds of inflation.

To summarize, the truncated assisted inflation model is a good
description for observable inflation in the limit that $\bar \mu \to
0$ (and the potential is quadratic) and $\bar \mu \to \infty$ (and the
potential is quartic).  For intermediate values $\O(10) \lesssim \bar
\mu \lesssim \langle \psi_i \rangle_{\rm in}$ (with $\langle \psi_i
\rangle_{\rm in}$ the average initial amplitude of the fields), the
truncated description may still apply but the parameters of the model
are different.

\subsection{M-flation: Extended SU(2) sector}
\label{s:multi_extended}

So far we have concentrated on the SU(2)-sector of M-flation with only
the trace field evolving during inflation, while all other fields are
spectators. M-flation differs from assisted inflation in that all
fields couple to each other, and it becomes much harder to determine
in full generality whether the truncated sector is an attractor
solution. In this subsection we will consider the simplest extension,
truncating to the ``extended SU(2)'' sector containing three fields,
valid for arbitrary $N_c$, while in the next subsection we consider
the full gauged $N_c=3$ model. As we will see, the SU(2)-sector is
generically not an attractor of this system.  We study how the
multi-field dynamics alters inflationary predictions.

The ``extended SU(2)'' sector is the simplest multi-field
generalization of the SU(2)-sector that allows for a consistent
truncation.  The three scalar fields $\phi_a$ are defined via $\Phi^i
= \phi_a^i J^i + \Pi^i$ with $\phi_a^i = \phi_a \delta_{ai}$:
\be 
\(
\begin{array}{c}
\Phi^1 \\ \Phi^2 \\ \Phi^3
\end{array}
\)
= 
\(
\begin{array}{c}
\phi_1 J^1+ \Pi^1 \\ \phi_2 J^2+\Pi^2 \\ \phi_3 J^3+\Pi^3
\end{array}
\).
\label{xphi}
\ee
The extended SU(2) sector corresponds to $\vec \Pi =0$. The Lagrangian
is $\L = \L_{\rm kin} - V$ with $\L_{\rm kin} = \frac12 \sum_a
(\partial_\mu \phi_a)^2$ and
\be 
V^{(0)}_X = 
\frac{\mu^2}{2} \sum_a \phi_a^2 +\sqrt{3}\k \phi_1 \phi_2 \phi_3 +
\frac{3\lambda}{4} ( \phi_1^2 \phi_2^2 + \phi_2^2 \phi_3^2 + \phi_1^2 \phi_3^2)
\label{xL}
\ee
where, as before, we rescaled the field to put the kinetic term in
canonical form \eref{rescale}, and absorbed factors of $d_c$ in the
couplings \eref{rescale2}.  The truncation to the extended SU(2)
sector is still consistent since the potential linear in $\vec \Pi$
\be V^{(1)}_X = \frac1{\sqrt{C_N}} {\rm Tr} \bigg\{ \Big( 6\lambda
\phi_a^2 \phi_b + \frac{2 \kappa}{\sqrt{3}} \phi_a \phi_b +
\frac{\mu^2}{2} \Big) \delta_{ai} \delta_{bj} J^i \Pi^i \bigg\}=0 \ee
vanishes as the trace ${\rm Tr}(J^i \chi^i)=0$ vanishes for each
SU(2)component separately (no summation over $i$ needed).  The
eigenvectors of \eref{ev} are still mass eigenstates, but now
$(l,m)$-dependent functions of $\phi_a$, that is the degeneracy with
$m$ is broken. We will assume that during inflation all spectator
masses are positive --- this is supported by $N_c=3$ model in
discussed in \S \ref{s:multi_N3}.

As a side remark, we can translate the above to the notation of \S
\ref{s:mflation}.  To do so define the trace field $\phi =
\sum_a\phi_a/\sqrt{3}$, where the normalization is chosen to get a
canonical kinetic term. Further identify in the traceless matrix $\vec
\Psi$ \eref{phi} the two independent models $\vec \psi_a$ with $a
=1,2$, that have components along $J^i$, that is ${\rm Tr}(\psi_a^i
J^i) \neq 0$ for some SU(2) components $i$.  The decomposition
\eref{phi} for the SU(2) sector can then be rewritten into the form
for the extended SU(2) sector \eref{xphi}: $ \Phi^i = \phi
J^i/\sqrt{d_c} + \Psi^i = (\phi + \psi^i_1+\psi^i_2) J^i/\sqrt{d_c} +
\vec \Pi $.  Choosing orthonormal modes $\Tr(\vec\psi_a \cdot \vec
\psi_b) = \delta_{ab}$, which ensures canonical kinetic terms for
$\phi, \vec \psi_a$, gives
\be
\vec \phi_1  = \frac{\vec \phi}{\sqrt{3}} +  \frac{\vec \psi_1}{\sqrt{2}} + 
\frac{\vec\psi_2}{\sqrt{6}}, \qquad
\vec \phi_2 = \frac{\vec \phi}{\sqrt{3}} - \frac{\vec\psi_1}{\sqrt{2}} + 
\frac{\vec\psi_2}{\sqrt{6}}, \qquad
\vec \phi_3 = \frac{\vec \phi}{\sqrt{3}} - 2 \frac{\vec\psi_2}{\sqrt{6}}
\label{phipsi}
\ee
where we defined $\vec \phi = \phi (1,1,1)$.  Substituting in the
potential \eref{xL} gives $V^{(0)}$ as before \eref{V0}, while the
potential quadratic in $\psi_a$ is
\be
V^{(2)} = \frac12 (\mu^2- \k \phi) \sum_{a=1,2} \psi_a^2 .
\ee
Comparing $V^{(2)}$ for the extended sector with the generic
expression $V^{(2)}$ in \eref{V2}, the two $\vec \psi_a$ modes can be
identified with the $l=1$ $\beta$-modes \footnote{To be precise, the
trace field $\phi = \alpha_{10}$, and $\phi_{1,2} = \pm {\rm
Re}(\beta_{12})/\sqrt{3} +(\alpha_{10}-\beta_{10}/\sqrt{2})/3$,
$\phi_3=(\alpha_{10}+ \sqrt{2}\beta_{10})/3$.  Here we labeled the
modes such that $\psi^3_{lm} \propto Y_{lm}$ --- see \cite{raamsdonk}
and appendix \ref{A:interactions} for more details. The expressions
for $\vec \psi_a$ can be found from \eref{phipsi}; since they are not mass
eigenstates they do not map to pure $\alpha$ or $\beta$ modes.}.

\subsubsection{Inflation}

\begin{figure}[t]
\centering 
\subfigure[Minimum $\phi_i^{\rm vac}$:
\newline 
\hspace*{.5cm}(blue, green) = $(0,\,\bar \mu)$\hspace*{3cm}]
{\includegraphics[scale=.42]{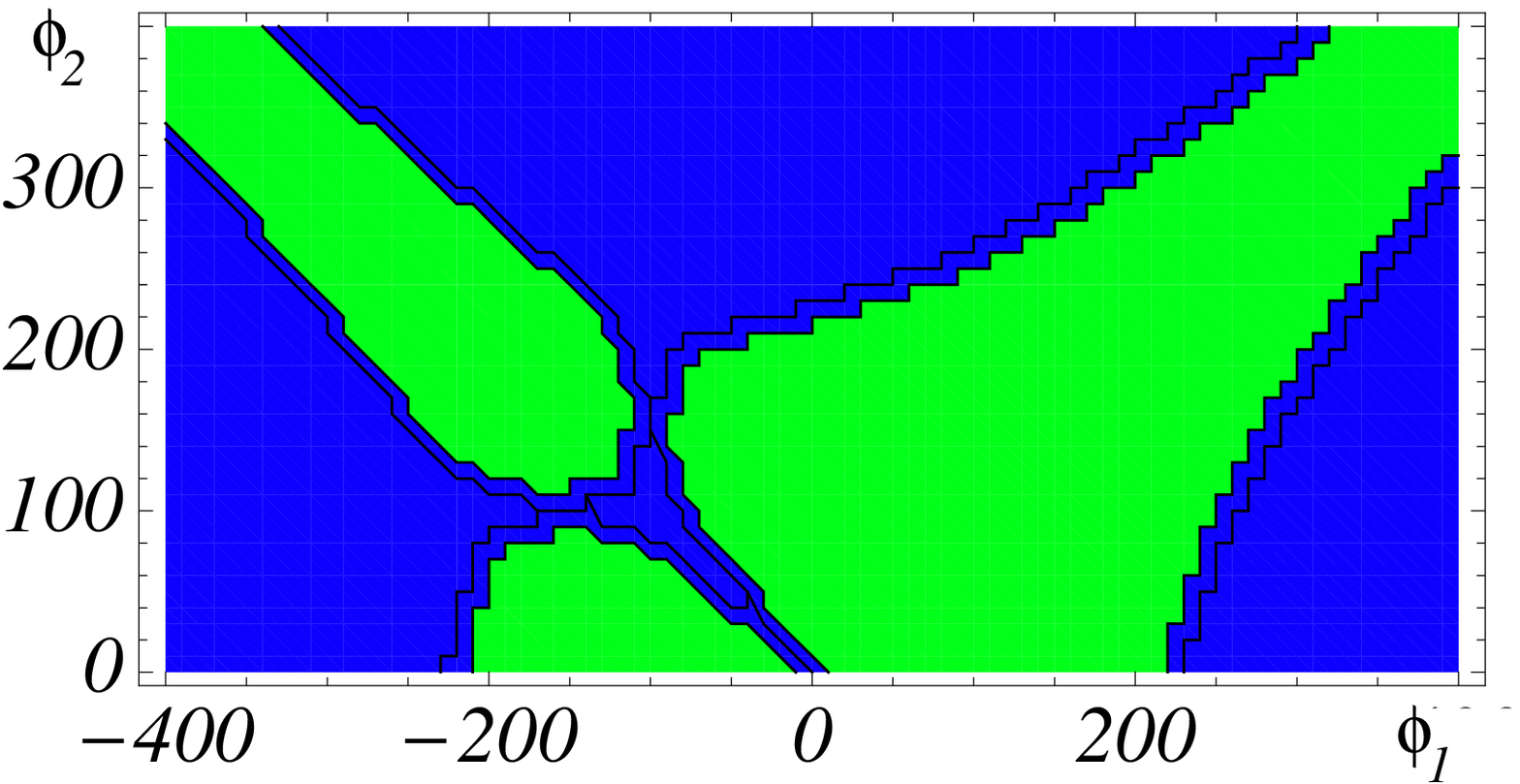}
\label{F:para}} 
\subfigure[Spectral index $n_s$:
\newline
\hspace*{.5cm}(blue, green, red) $\approx (<0.967,\, 0.964,\,>0.97)$
 ]{ \includegraphics[scale=.5]{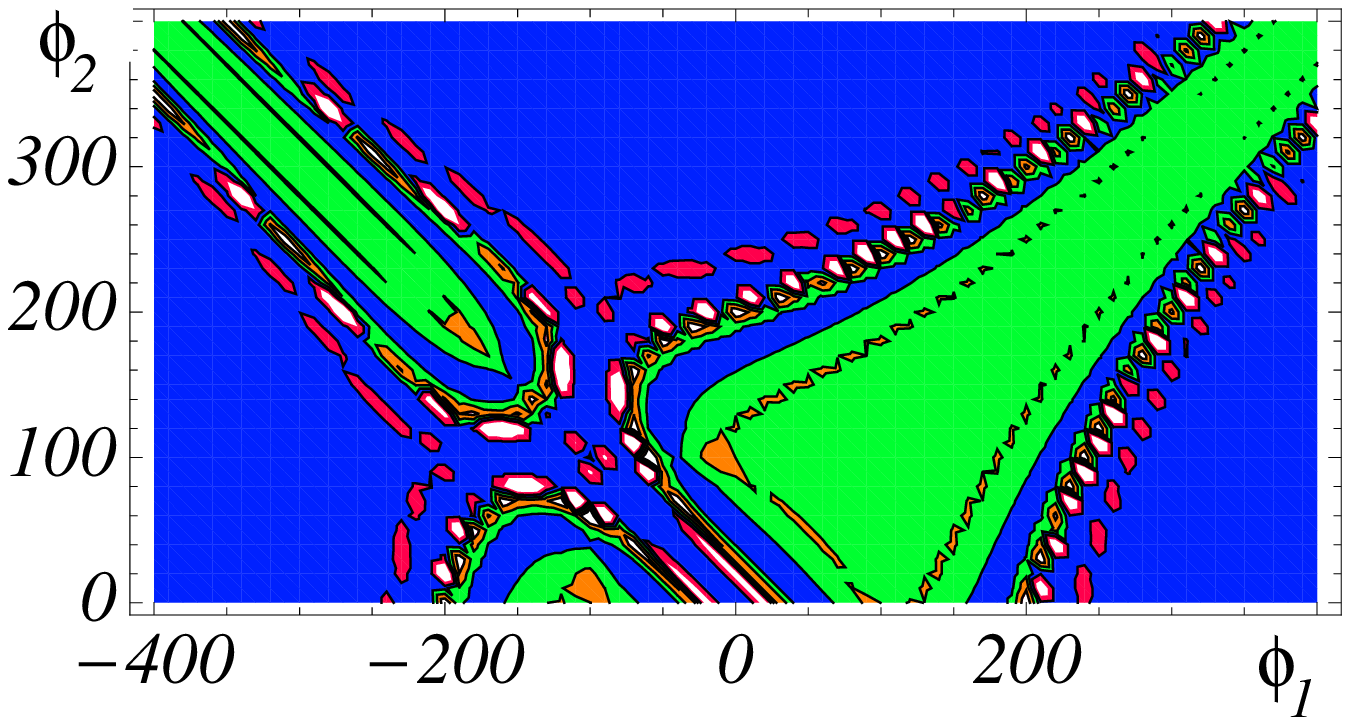}
\label{F:parb}
}
\label{F:par}
\caption[]{M-flation in extended SU(2) sector: Minimum and spectral
for different initial values $\phi_1,\phi_2$, with $\bar \mu =40$ and
$\phi_3 = 10^2$ kept fixed.}
%
\subfigure[Case 1a: $(\phi_1)_{\rm in} \gtrsim 5 \bar \mu$, 
a typical example of the blue region in Fig. 2a.]{
\includegraphics[scale=.6]{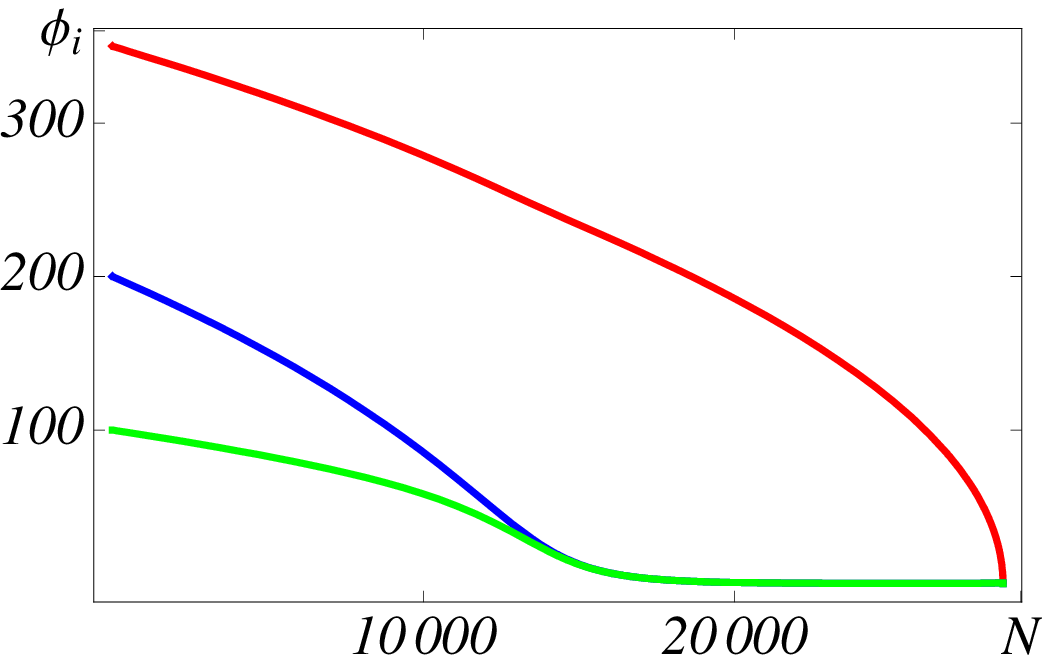}
\label{F:evola}
}\hspace*{0.5cm}
\subfigure[Case 2a: $(\phi_i)_{\rm in} \lesssim  5 \bar \mu$,
a typical example of the green region in Fig. 2a.]{
\includegraphics[scale=.6]{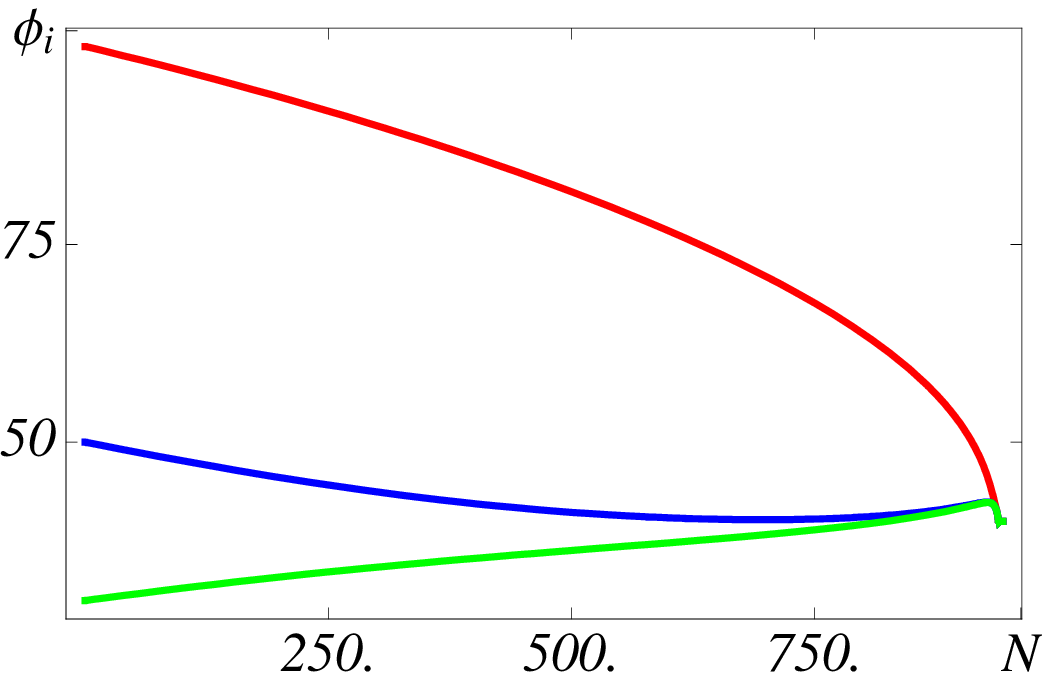}
\label{F:evolb}
}
\label{F:evol}
\caption[]{M-flation in extended SU(2) sector: evolution of the fields
$\phi_1,\phi_2,\phi_3$ as a function of the number of e-folds $\N$ for
$\bar \mu =40$ and different initial field values.}
\end{figure}

We will discuss inflation in the extended SU(2) sector in terms of the
$\phi_a$ fields, rather than using the trace field and $\psi_a$ or the
$\alpha$ and $\beta$ modes, as the former enters the potential in a
symmetric form \eref{xL}. As before we take $\kappa = -3
\sqrt{\lambda/2} \mu$ \eref{kappa}. In the limit that all fields are
equal $\phi_i =\phi$, the potential reduces to the ``symmetry
breaking''-potential for the trace field \eref{Vsymmb}; we will refer
to this single field model discussed in \S \ref{s:traceflation} as
``trace inflation''. The generic 3-field potential also has two
minima, one with all fields at the origin $\phi_i =0$ and one symmetry
breaking minimum at $\phi_i = \bar \mu$.  We numerically evolved the
three fields through inflation, starting with some given initial
conditions, and calculated the spectrum of density perturbations.
More details on the procedure can be found in Appendix
\ref{A:perturbations}

First consider the two limiting cases $\bar \mu \to \infty$ and $\bar
\mu \to 0$ for which the potential reduces to a quadratic and quartic
potential respectively.  In the quadratic case, the fields decouple,
and inflation is effectively single field. However, in the quartic
case the trace field is not a late time attractor. Unlike the
uncoupled case of assisted inflation, where the field with the largest
amplitude starts rolling first and picks up all the others along the
way, now it is the field with the smallest amplitude that has the
steepest potential and goes first. For definiteness, we order
$|\phi_1|_\ini > |\phi_2|_\ini > |\phi_3|_\ini$, with the subscript
``$\ini$'' denoting the initial field values.  Then $\phi_3$ will roll
down first; during the last 60 e-folds, only $\phi_2$ is rolling while
$\phi_1$ is still frozen at a large amplitude. It is model dependent
whether such a set-up gives a good post-inflationary phenomenology.
If a mass term kicks in before $\phi_1$ decays, which likely happens
for $\bar \mu$ small but not absolutely zero, there will be an
additional period of quadratic inflation driven by $\phi_1$.

Consider now intermediate $\bar \mu$ values.  Just as for the quartic
potential, trace inflation is not an attractor solution. But are there
other late time attractors?  Yes. The parameter space of initial
conditions can be divided into two regions, corresponding to whether
the system relaxes to the symmetry breaking minimum $\phi_i = \bar
\mu$ or to the symmetry restoring minimum $\phi_i =0$.  The mass of
each field is roughly inversely proportional to its amplitude.
Consequently, the field with the largest initial field value is light
and remains initially frozen, while the two other slightly heavier
fields roll towards their minimum. Depending on initial conditions, in
particular on the amplitude of the lightest field, the minimum is
either the origin or the symmetry breaking minimum. The inflation
phenomenology is fully determined by the vacuum the system ends up
in. We order again $|\phi_1|_\ini > |\phi_2|_\ini >
|\phi_3|_\ini$. Further we introduce the notation $\Delta_{ij} =
(|\phi_i|_\ini- |\phi_j|_\ini)/ |\phi_i|_\ini$ which measures the degree
of degeneracy of the initial field values. Parameter space divides
into two regions depending on $|\phi_1|_\ini$ and $\Delta_{ij}$. To be
precise
\begin{enumerate}
\item
The system relaxes to the $\phi_i =0$ vacuum, if:
\begin{enumerate}
\item
$|\phi_1|_\ini \gtrsim 5 \bar \mu$ large and $\delta_{12} > \O(10) \mu^2$
non-degenerate. This covers most of parameter space for large initial
$\phi_1$ amplitude.
\item 
$|\phi_1|_\ini \lesssim 5 \bar \mu$ small and $\delta_{23} < \O(10) \mu^2$
degenerate with $\phi_2 = - {\rm sign}(\mu)\phi_3$. This is a small
valley in parameter space for small initial $\phi_1$ amplitude.
\end{enumerate}
\item
The system relaxes to the $\phi_i = \bar \mu$ vacuum, if:
\begin{enumerate}
\item
$|\phi_1|_\ini \lesssim 5 \bar \mu$ small and $\delta_{12} > \O(10) \mu^2$
non-degenerate. This covers most of parameter space for small initial
$\phi_1$ amplitude.
\item 
$|\phi_1|_\ini \gtrsim 5 \bar \mu$ large and $\delta_{12} < \O(10) \mu^2$
degenerate. This is a an increasingly small valley in parameter space
for larger initial $\phi_1$ amplitude.
\end{enumerate}
\end{enumerate}
Note that 1(b) and 2(a) include as a special case trace inflation with
all fields equal $\phi_i = \phi$. These (numerical) results are
illustrated in Fig.~\ref{F:para}, which shows the the minimum as a
function of $\{(\phi_1)_\ini,(\phi_2)_\ini\}$, with $(\phi_3)_\ini
=10^2$ fixed and $\bar \mu =40$.  Fig.~\ref{F:parb} shows the spectral
index which, except for the boundary regions, is fully correlated with
the vacuum structure.
We will now discuss these results in more detail.

Consider first the $(\phi_i =0)$-minimum of case 1.  For these
initial conditions the two relatively heavy fields $\phi_2,\phi_3$
roll towards the origin, as shown in Fig.~\ref{F:evola}. With two of
the three fields sitting at the origin the potential takes on the
simple form $V|_{\phi_2 =\phi_3=0} = (1/2)\lambda \bar \mu^2
\phi_1^2$. During the last 60 e-folds only the adiabatic inflaton mode
$\phi_1$ rolls in a quadratic potential, giving rise to the usual
single field results.  Even though $\phi_2,\phi_3$ may be light during
inflation and fluctuate, there is no subsequent conversion of these
isocurvature modes into curvature modes during inflation, as both
fields are trapped at the origin. In other words, observationally this
set-up is indistinguishable from a single field chaotic inflation
model with a quadratic potential.

Now the symmetry breaking minimum of case 2. The two relatively heavy
fields $\phi_2,\phi_3$ roll towards their instantaneous minimum which
is close to $\phi_i =\bar \mu$, as shown in Fig.~\ref{F:evolb}. During
the last 60-e-folds all three fields roll towards the symmetry
breaking minimum, but since $\phi_2,\phi_3$ are already close, to a
good approximation the inflaton can be identified with the initially
large field $\phi_1$ and multi-field effects are small --- but not
necessarily unobservable small. The effective inflaton potential is
quadratic plus corrections; these corrections cancel to first order in
the expression for the spectral index, but can be of order $\O(10\%)$
for the tensor to scalar ratio.  This is similar to what we saw in
trace inflation \eref{sr_corrections}.

To summarize, for large initial field values (and a degenerate valley
at small amplitudes) quadratic inflation is an attractor solution.
Inflation is effectively single field during the last 60 e-folds, and
the effect of the isocurvature perturbations on the observables is
negligible.  For smaller initial values (and a degenerate valley at
large amplitudes) the corrections from quadratic inflation, both in
$n_s$ and $r$, are appreciable.  Inflation is approximately single
field, but observable 10\% deviations from single field inflation are
possible.  

\subsection{The gauged $N_c =3$ model}
\label{s:multi_N3}

In this section we discuss the gauged $N_c =3$ model, which has 4
$\alpha$- and 15 $\beta$-modes for a total of 19 fields $\phi_i$. The
8 zero modes are eaten by the gauge bosons, which are heavy and can be
integrated out.  Although 19 fields is a small number compared to the
large number of fields usually considered in assisted inflation, we
think it is nevertheless large enough to get a good qualitative
description of the dynamics of such multi-multi-field models. The
interaction terms between the different fields is discussed in more
detail in appendix \ref{A:interactions}. The potential is too
complicated to analyze analytically, and we mainly rely on our
numerical results in this section.

Most of our results are just an extrapolation of what is already
observed in the 3-field model of the extended SU(2) sector just
discussed in \S \ref{s:multi_extended}.  First of all, trace inflation
is not an attractor solution. The trace field can be identified with
the $\alpha_{10}$-mode.  Trace inflation would be an attractor if all
fields would settle in their minimum, with only $\alpha_{10}$ rolling
during the last 60 e-folds. What is seen, instead, is that all fields
roll during the last e-folds. Figs.~\ref{F:M3}, \ref{F:M2},
\ref{F:M1} show the evolution of all 19 fields as a function of the
number of e-folds for different parameters and initial conditions.

In the limit $\langle \phi_i \rangle_{\rm in} \ll \bar \mu$, with as
before $\langle \phi_i \rangle_{\rm in}$ the average initial field
value, the quadratic terms dominate the potential. By construction the
quadratic terms are diagonal, and thus the fields are
uncoupled. Hence, in this limit we recover assisted quadratic
inflation. By making a rotation in field space, one can always set all
field values to zero, except for one field which has amplitude
$\phi_{\rm eff}^2 = \sum_i \phi_i^2$.  This procedure makes it
manifest that inflation can be described by a single field model, with
the usual observables for quadratic inflation and without any
multi-field corrections. This is illustrated in Fig.~\ref{F:M3}. The
relative field amplitudes remain constant over time, which is the
tell-tale sign of degenerate quadratic potentials.

In the opposite limit $\langle \phi_i \rangle_{\rm in} \gg \bar \mu$
the quartic terms initially dominate the potential.  All fields
plummet down the potential, along the steepest direction set by the
quartic terms, and within a relatively short time the fields lose
much of their amplitude. As a consequence, the quartic and cubic terms
become negligibly small, and the rest of the evolution is dominated by
the quadratic terms.  In particular this means that the last 60
observable e-folds of inflation are well described by an assisted
quadratic inflation model.  This is shown in Fig.~\ref{F:M2}.

We see that whatever the initial field values, single field quadratic
inflation is a late time attractor of the system.  The only possible
exception is when $\langle \phi_i \rangle_{\rm in} \sim \bar \mu$ ---
also in the extended SU(2) sectors we saw the largest deviation from
single field inflation in this parameter regime. But now a new element
kicks in.  Whereas in the 3-field model of \S \ref{s:multi_extended}
the system always evolves towards the stable minimum with $V_{\rm min}
=0$, for the multi-field $N_c =3$ model with 19 fields this is not the
case. In the two extreme limits $\langle \phi_i \rangle_{\rm in} \ll
\bar \mu$ and $\langle \phi_i \rangle_{\rm in} \gg \bar \mu$ the
potential is at all times dominated by either the quadratic or the
quartic terms and this problem does not arise, the fields evolve
towards $\phi_i = 0$.  However for the intermediate $\bar \mu$ values
under consideration this is not the case; the cubic terms may at some
point during the evolution become important causing the system to
overshoot and run off to some minimum with $V <0$, ruining
inflation. This is illustrated in Fig.~\ref{F:M1}.
What happens is that initially the quartic terms dominate
but not by much. Instead of the quartic and cubic term immediately
plummeting down to negligible values, as happens in the $\langle
\phi_i \rangle_{\rm in} \gg \bar \mu$ limit, they only decrease by a
moderate amount, and the cubic terms get a chance to dominate the
potential later on.

\begin{figure}[ht]
\centering
\includegraphics[scale=.8]{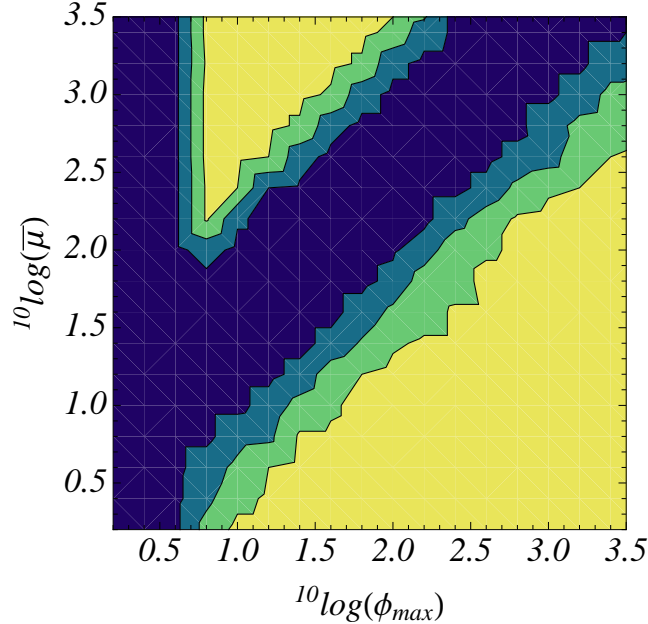}
\caption[]{Gauged M-flation ($N=19$, $N_c=3$): Chance $P$ of
succesfull inflation for random inital conditions taken from a flat
distribution for $(\phi_i)_\ini^2$, with $|(\phi_i)_\ini|< \phi_{\rm
max}$ and different values of $\bar \mu$. $P < 0.1$ in the blue region
(no inflation), $P < 0.5$ in dark green, $P <0.9$ in light green, and
$P=1$ in the yellow region (succesful inflation).\label{F:Mmin} }
\end{figure}

Fig.~4 shows the chance of successful inflation as a function of $\bar
\mu$ and $\phi_{\rm max}$, where we choose random initial conditons
for all fields from a flat distribution for $(\phi_i)_\ini^2$ and
$(\phi_i)_\ini \in[-\phi_{\rm max}, \phi_{\rm max}]$. For too small
initial amplitudes $\langle \phi_i \rangle_\ini$, the initial slow
roll parameters are too large to generate at least 60 e-folds of
inflation. The other region where inflation does not work is for $\bar
\mu \sim \langle \phi_i \rangle_\ini$, since, as explained above, the
system evolves to the wrong minimum. In the whole parameter region
where inflation takes place (except very near the boundary), the last
60 e-folds correspond to quadratic assisted inflation with $n_s
\approx 0.67$, and the deviations from single field inflation are
unobservably small. 

To summarize, trace inflation is not an attractor solution. Starting
with random initial amplitudes for all fields, either the system
evolves towards an Anti-de Sitter vacuum and inflation does not work
(for intermediate $\langle \phi_i \rangle_{\rm in} \sim \bar \mu$), or
quadratic assisted inflation with all masses degenerate is a late time
attractor. In the latter case, the inflationary results are as in
usual assisted inflation.  In particular the fields and and quartic
coupling scale by the number of fields $N$ \eref{scale_ass}, rather
than a power $N^{3/2}$ as found in truncated M-flation
\eref{rescale2}.

\section{Conclusions}
\label{s:conclusions}

In this paper we discussed the role of the spectator fields in M-flation.  

In M-flation the inflaton fields are three $N_c \times N_c$
non-abelian hermitian matrices, corresponding to $N=3N_c^2$ degrees of
freedom.  The model can be consistently truncated to an effectively
single field inflation model, with all spectator fields fixed at the
origin. This set-up can be viewed as an implementation of assisted
inflation, in the sense that the effective inflaton field and its
effective couplings are scaled up/down by the many fields in the
theory. There is however an important difference: whereas in typical
assisted models all fields are decoupled, the spectator fields of
M-flation couple to each other.  This leads to an enhanced assisted
effect, and the fields and couplings scale stronger with the number of
fields. However, this goes against the usual lore that cross-couplings
tend to destroy the assisted inflation.  How does this work for
M-flation?

It does not work. Inflation in the truncated sector, dubbed trace
inflation, is not a late time attractor of the system.  Instead,
starting with random initial amplitudes for all fields, the late-time
attractor is quadratic assisted inflation with degenerate masses.  All
fields are decoupled, rolling down the potential simultaneously.  The
usual results for assisted inflation apply, in particular the usual
scaling with the fields.

We also looked at the UV behavior of the theory.  In models with many
species the effective quantum gravity scale is lowered $\Lambda =
\mpl/\sqrt{N}$.  Demanding that the energy density during inflation is
below the quantum gravity scale bounds the number of fields in the
theory $N \lesssim 10^4$.  The assisted effect is thus limited. For
example the effective quartic self-coupling $\lambda = \lambda_0/N$ is
at most scaled down by a factor $10^4$, and the coupling appearing in
the original Lagrangian still has to be tuned small $\lambda_0
\lesssim 10^{-10}$.  Viewing the inflaton model as a low-energy
effective field theory, the control over the unknown physics at the UV
scale is not enhanced with respect to usual single field chaotic
inflation.  The UV behavior in trace inflation is even worse.  But
this result is moot, as we have seen that trace inflation is not a
late time attractor, and can only occur for fine-tuned initial
conditions; for generic initial conditions M-flation evolves towards
assisted inflation.

\section*{Acknowledgment}
The author thanks Damien George and Sander Mooij for usefull comments,
and the Dutch Science Organization (NWO) for financial support in the
form of a VIDI grant.


\appendix


\section{Perturbations in multi-field inflation}
\label{A:perturbations}

In single field inflation there is only one scalar perturbation, the
adiabatic perturbation, corresponding to fluctuations along the
direction of the background trajectory. In multi-field inflation there
are isocurvature perturbations as well, fluctuations orthogonal to the
background trajectory. If the trajectory in field space changes
direction, isocurvature perturbations source adiabatic perturbations
\cite{gordon}.  In the slow roll limit, the rate of change of
direction is suppressed.  Nevertheless, the transfer depends on the
total change of direction, which is an integration over time, and the
cumulative effect can still be significant. This allows for a direct
probe of the multi-field nature of the model.

Consider a system of $N$ canonically normalized fields $\{\phi_i\}$,
with $i$ numbering the light fields.  It is useful to go to a special
basis in field space \cite{gordon}.  Introduce a set of orthonormal
basis $\{\vec{e}_n\}$.  Align the first vector along the direction of
the background trajectory $e^i_\sigma = \dot \phi^i/\dot \sigma$ with
$\sigma = e^i_\sigma \phi^i$ the adiabatic background direction, and
$\dot \sigma = \sum_i \dot \phi_i^2$. All other $j=1,..,N-1$ vectors
$\vec{e}_{{\bar s}_j}$ are orthonormal to the adiabatic mode. Note
that the vectors $\vec e_n$ may change with time. The slow roll
parameters generalize to
\be
\eps_n
= \frac12 \( \frac{e_n^i V_i}{V} \)^2,
\qquad
\eta_{nm} = \frac{V_{ij} e^i_n e^j_m }{V},
\ee
where we assumed canonical kinetic terms for all fields, and defined
$V_i = \partial_{\phi_i} V$ etc. They can be either calculated in the
new basis with $n=\{\sigma,s,\bar s_j\}$, or in the original basis
$\{\phi^i\}$ using $\e^i_j = \delta^i_j$.  Inflation requires $\eps
\equiv \sum \eps_n \ll 1$; to get sufficient e-foldings $\eta_{\sigma
\sigma} \ll 1$ as well.  

The number of e-folds to the end of inflation along the unperturbed
path is
\be
\N 
= \int_{\sigma_*}^{\sigma_e} \frac{H} {\dot{\sigma}} \dd \sigma.
\ee
with $\sigma_e$ the field value at the end of inflation when $\eps =
1$.  The fluctuations in the number of e-folds when $\phi_i \to \phi_i
+ \delta \phi_i$ is \cite{gordon,tye}
\be
\delta \N = -\frac{H}{\dot \sigma} (\partial_i \sigma) \delta \phi^i \bigg|_*
+ \frac{H}{\dot \sigma} (\partial_i \sigma_e) \delta \phi^i \bigg|_e
-\int_{t^*}^{t_e} \dd t \frac{2H}{\dot \sigma} 
\dot e^i_\sigma \delta \phi^i.
\ee
The first term is the adiabatic perturbation produced at horizon
exit. In single field inflation this would be the only contribution
and $\dot \zeta =0$ freezes as soon as $\delta \phi$ freezes, which is
just after horizon exit. The second term contributes when the
hyper-surface of the end of inflation is not orthogonal to the
background inflation path. We will assume this term to be small. For
the 3-field model of \S\ref{s:multi_extended} we verified this
explicitly, but it remains te be checked in full generality.  Finally
the last term captures the continuous sourcing of the adiabatic
perturbations by the isocurvature perturbations, as the latter change
the length of and speed along the inflationary trajectory. 

For light modes the adiabatic and curvature perturbations are $\delta
\sigma,\,\delta s_j \sim {H}/({2\pi})$. Expanding to first order
$\delta \N = \N_i \delta \phi^i$ we can read off
\be
\N_\sigma = -\frac{1}{ \sqrt{2\eps_*}},\qquad
\N_s = 
-2 \int_{\N_*}^{\N_e} \dd \N \frac{2}{\sqrt{2\eps}} 
({\vec e_\sigma\, '} \cdot {\vec e_{s_*}}),
\label{Ni}
\ee
where prime denotes derivative w.r.t.~to the number of e-folds $d\N =
H dt$, and we used $\sigma' = \sqrt{2\eps}$. We take $\N_e - \N_* =
60$ e-folds of observable inflation. In slow roll inflation the
integrand of $\N_s$ is suppressed by the smallness of the slow roll
parameters.  Nevertheless, since the integration is over the last 60
e-folds, the accumulated effect can be order one. The curvature
perturbation at the end of inflation is then \cite{tye,bartolo}
\be
\P_\zeta = \frac{H^2}{4\pi^2}(\N_\sigma^2+ \sum_j \N_{s_j}^2)
= \frac{P_\zeta|_*}{ \sin^{2} \Delta},
\ee
with $P_\zeta|_* = ({H^2}/{4\pi^2}) \N_\sigma^2$ and $\Delta$ the
correlation angle between the curvature and isocurvature modes which
parametrizes the growth of the curvature perturbation after horizon
exit, fed by the isocurvature perturbations.  Explicitly
$
\sin^2 \Delta =  {\N_\sigma^2}/{(\N_\sigma^2+ N_s^2)}.
$
In the limit of effectively single field inflation, and no sourcing of
adiabatic perturbations, $\sin^2 \Delta = 1$.  

The tensor perturbations remain frozen $\P_T = \P_T |_*$, and the
consistency condition for the tensor-scalar amplitudes becomes
\cite{bartolo,wands,polarski,sasaki,bellido}
\be 
r = \frac{\P_T}{\P_\zeta} = - 8 n_T \sin^2 \Delta =16 \eps \sin^2
\Delta, 
\label{sinD}
\ee
with $n_T$ the spectral index of the tensor signal.  Hence, $\sin^2
\Delta$ measures the deviation from single field inflation which can
be probed experimentally if a gravitational wave signal is detected.
The calculation of $\sin^2 \Delta$ requires the integration of all
field perturbations \cite{vanTent}, which is beyond the scope of this
paper.  The values of the power spectrum, spectral index, and
scalar-to-tensor rate quoted throughout the paper correspond to the
contribution of the adiabatic mode only, with the isocurvature modes
negelected. This is a very good approximation for most of parameter
space (whenever the attractor solution is reached, inflation is
effectively single field).  


Numerically, we solve the equations of motion as formulated in (5.1)
of \cite{Bracetrack}; (5.2) of the same reference gives the expression
for the adiabatic power spectrum denoted $\P_{\zeta_*}$ here.  The adiabatic
spectral index is $(n_s-1) = \dd \ln \P_{\zeta_*}/\dd \N$.


\section{Interaction terms}
\label{A:interactions}

In this appendix we will concentrate on the gauged model, and only
consider the interaction terms for the $\alpha$ and $\beta$ modes. The
trace field can be identified with the $\alpha_{10}$ mode; we treat it
here on equal footing with all other fields.  We just summarize the
main results; more details and derivations can be found in
\cite{raamsdonk}.

Introduce the notation $\rho,\sigma,\theta,\mu = \{klm\}$ with
$k=\alpha,\beta$ for the alpha and beta-modes respectively, and
$\{lm\}$ the angular momenta under SU(2). Further $Y_{\rho_z} =
Y_{lm}$ and $Y_{\rho_\pm} = Y_{l(m\pm 1)}$, with $Y_{lm}$ the
irreducible spin $lm$ representation of SU(2), normalized to
\be
{\rm Tr}(Y_{l'm'}^\dagger Y_{lm}) = \delta_{l'l} \delta_{m'm}.
\label{norm}
\ee
Defining $\Phi^\pm = \Phi^1 \pm i \Phi^2$, we can expand the three
$N_c \times N_c$ matrices
\be
\vec \Phi = 
\left(
\begin{array}{c} 
\Phi^+ \\ \Phi^- \\ \Phi^3
\end{array}
\right)
=
\sum_\rho {x_\rho}
\left(
\begin{array}{c} 
f_+(\rho) Y_{\rho_+} \\ f_-(\rho) Y_{\rho_-}  
\\ f_3(\rho) Y_{\rho_3} 
\end{array}
\right),
\label{modes}
\ee
with
\ba 
\{f_\pm(\rho) , f_3(\rho)\}_{\rho=(\alpha,l,m)}
&=&\frac{1}{\sqrt{l(2l+1)}} \{ \sqrt{(l\pm m)(l+1 \pm m)},\sqrt{(l+m)(l-m)}\} ,
\label{modefunctions} \\
\{f_\pm(\rho), f_3(\rho)\}_{\rho = (\beta,l,m)}
&=&\frac{1}{\sqrt{(l+1)(2l+1)}} \{ \sqrt{(l\mp m)(l+1 \mp m)},\sqrt{(l+1+m)(l+1-m)}\} ,
\nn
\ea
for $\alpha$ and $\beta$ modes respectively.

\paragraph{Kinetic and quadratic terms}
With these definitions the kinetic terms are canonical for the
$x_\rho$ modes:
\ba
\L_{\rm kin} &=& \frac12 {\rm Tr}\( \partial_\mu \Phi^i \partial^\mu \Phi^i\)
\nn \\
&=& \frac12 \partial_\mu x_{klm} \partial^\mu x_{k'lm}^* \left[ 
- \frac12(f_+(klm) f_-(k'l-m) + f_-(klm) f_+(k'l-m))+ f_z(klm) f_z(k'l-m)
\right] 
\nn \\
&=&\frac12  |\partial x_{klm}|^2.
\ea
Here we used $Y_{lm} = (-1)^m Y^\dagger_{l-m}$, and $x^*_{lm} = (-1)^m
x_{l-m}$, which both follow from the reality condition for hermitian
matrices.  Further we used that for any vector $\vec x$ we have
$\sum_i x^i x^i = (x^1)^2 + (x^2)^2 + (x^3)^2 = 1/2(x^+ x^-+x^- x^+) +
(x^3)^2$, with $x^\pm = x^1 \pm i x^2$.  We can express the Lagrangian
in terms of manifestly real fields by writing
\be
{\rm Re}(x_{lm}) = \frac12\(x_{lm} + (-1)^m x_{l-m}\),
\qquad
{\rm Im}(x_{lm}) = \frac12\(x_{lm} - (-1)^m x_{l-m}\).
\ee
Similarly to the kinetic terms, the quadratic terms in the potential
are diagonal
\be
V^{(2)} = \frac12 {\rm Tr}\( \Phi^i \Phi^i\)
=\frac12  |\partial x_{klm}|^2.
\ee

\paragraph{Cubic interactions}

The cubic interactions are
\ba
V^{(3)} &=& - \frac{i\kappa_0}{6}{\rm Tr}(\eps_{ijk}[\Phi^i,\Phi^j]\Phi^k)
=- \frac{\kappa_0}{6}{\rm Tr}([\Phi^-,\Phi^+]\Phi^3 + {\rm cycl.})
\nn \\
&=&- \frac{\kappa_0}{2 N_c^{3/2}}
x_\rho x_\sigma x_\theta f_3(\rho) f_-(\sigma) f_+(\theta) 
b(\sigma_-,\theta_+,\rho_3) ,
\ea
with ``cycl.'' meaning cyclic permutations in $\{-,+,3\}$.  In the 2nd
step we used $\eps_{-+3} = 1/(2i)$. To get the last line we used that
$[Y_\rho,Y_\sigma]=N_c^{-3/2} b(\rho,\sigma,\theta) Y^\dagger_\theta$
and the normalization conditions \eref{norm}. Here
\ba
b(\rho,\sigma,\theta) &=& \frac12 \(1 - (-1)^{l_\rho + l_\sigma + l_\theta}\)
\nn \\
&& (-1)^{N_c} N_c^{3/2} \sqrt{(2l_\rho+1)(2l_\sigma+1)(2l_\theta+1)}
\left( 
\begin{array}{ccc}
l_\rho & l_\sigma & l_\theta \\
m_\rho & m_\sigma & m_\theta
\end{array}
\right)
\left\{
\begin{array}{ccc}
l_\rho & l_\sigma & l_\theta \\
\Theta & \Theta &  \Theta
\end{array}
\right\} ,
\ea
with $\Theta =(N_c-1)/2$ and $(...)$ and $\{...\}$ the Wigner 3j and 6j
symbol. The function $b$ is non-zero only if
\ba
(1)& \hspace{-.5cm} l_\rho+ l_\sigma+ l_\theta = {\rm odd}, \qquad 
&(3)\;\; m_\rho+ m_\sigma+ m_\theta =0 ,\nn \\
(2)&\; |l_\sigma - l_\theta| \leq l_\rho \leq l_\sigma + l_\theta, \qquad
&(4)\; \;|l_i - \Theta| \leq \Theta \leq l_i + \Theta.
\label{b_cond}
\ea
Moreover $b$ is symmetric under cyclic permutations of $\{\rho,
\sigma, \theta\}$ and picks up a factor $(-1)^{^{\sum_i l_i}}$ under
$\rho \leftrightarrow \sigma$ or $m_i \leftrightarrow - m_i$.

\paragraph{Quartic interactions}

The quartic Lagrangian is
\ba
V^{(4)} &=&  
\frac{-\lambda_0}{4} {\rm Tr}\big( 
[\Phi_\rho^i,\Phi_\sigma^j ][\Phi_\rho^i,\Phi_\sigma^j ]
\big)
\nn \\
&=& -\frac{\lambda_0}{4}  x_\rho x_\sigma x_\theta x_\nu
f_i(\rho) f_j(\sigma) f_i (\theta) f_j(\nu)
{\rm Tr}\bigg([Y_{\rho_i},Y_{\sigma_j}][Y_{\theta_i},Y_{\nu_j}]
\bigg)
\nn\\
&=& -\frac{\lambda_0}{4N_c^3}  x_\rho x_\sigma x_\theta x_\nu 
\bigg(2 A[+,3,-,3]-\frac12A [-,+,-,+]
\bigg),
\ea
with
\be
A[i,j,k,l]= \sum_{lm} (-1)^m f_i(\rho) f_j(\sigma) f_k(\theta) f_l(\nu)
b\Big(\rho_i,\sigma_j,{lm}\Big) b\Big(\theta_k,\nu_l,\{l\! - \! m\}\Big).
\ee

\subsection{$N_c =2$ potential}

We can used the above expressions to find the explicit form of the
potential.  As an example we give the potential for the $N_c =2$
gauged model $V = V^{(2)}+ V^{(3)}+V^{(4)}$:
\ba
V  &=& \frac14 \lambda \bar \mu^2 \sum_i x_i^2 \nn \\
 &+& \frac12 \lambda \bar \mu \Big(
-x_1^3+\frac{3}{2} x_1 x_5^2-\frac{1}{\sqrt{2}}x_5^3 +3 x_1 x_6^2
-\frac{3}{\sqrt{2}} x_5 x_6^2 +3 x_1 x_7^2-\frac{3}{\sqrt{2}} x_5 x_7^2
-3\sqrt{3} x_6^2 x_8 \nn \\
&& \hspace{1.1cm} + 3 \sqrt{3} x_7^2 x_8+3 x_1 x_8^2+3 \sqrt{2} x_5 x_8^2
-6 \sqrt{3} x_6 x_7 x_9 + 3 x_1 x_9^2+3 \sqrt{2} x_5 x_9^2
\Big) \nn \\
 &+& \frac14 \lambda  \Big(
\frac{x_1^4}{2}-\frac{x_1 x_5^3}{\sqrt{2}} +\frac{3}{8} x_5^4
-\frac{3}{\sqrt{2}} x_1 x_5 x_6^2+\frac{3}{2} x_5^2 x_6^2
+\frac{3}{2}x_6^4 - \frac{3}{\sqrt{2}} x_1 x_5 x_7^2
+\frac{3}{2} x_5^2 x_7^2 + 3 x_6^2 x_7^2 \nn \\
&&\hspace{.7cm} + \frac{3}{2} x_7^4
-3 \sqrt{3} x_1 x_6^2 x_8+3 \sqrt{3} x_1 x_7^2 x_8 + 
3 \sqrt{2} x_1 x_5 x_8^2 + \frac{3}{2} x_5^2 x_8^2 + 3 x_6^2 x_8^2
+3 x_7^2 x_8^2\nn \\ 
&& \hspace{.7cm} + \frac{3}{2} x_8^4 - 6 \sqrt{3} x_1 x_6 x_7 x_9
+ 3 \sqrt{2} x_1 x_5 x_9^2 + \frac{3}{2} x_5^2 x_9^2 + 3 x_6^2 x_9^2
+ 3 x_7^2 x_9^2 + 3 x_8^2 x_9^2 +\frac{3}{2} x_9^4
\Big). \nn \\
\ea
Here we labeled the fields
\ba
&&\alpha_{10}= x_1, \;\beta_{00}= x_2, \;
\beta_{01}= x_3+i x_4, \; \beta_{0-1}= -x_3+i x_4,  \;
\beta_{10}= x_5, \;
\nn\\ 
&& 
\beta_{11}= x_6+i x_7, \;\beta_{1-1}= -x_6+i x_7,  \;
\beta_{12}= x_8+i x_9, \;\beta_{1-2}= x_8-i x_9.
\ea
Likewise the potential for the $N_c =3$ gauged model can be derived;
this is used in \S\ref{s:multi_N3}. The expression is too long to
reproduce here.

\end{document}